\documentclass[aps,twocolumn,superscriptaddress,prl]{revtex4-2}
\usepackage[utf8]{inputenc}
\usepackage{amsmath, amssymb, graphicx, color, subfigure, enumitem}
\usepackage[bookmarks=false]{hyperref}
\usepackage[usenames,dvipsnames]{xcolor}
\usepackage[super]{nth}
\usepackage[normalem]{ulem}

\def\CVS{CsV$_3$Sb$_5$}
\def\AVS{AV$_3$Sb$_5$}

\def\be{\begin{equation}}
\def\ee{\end{equation}}
\def\ba{\begin{eqnarray}}
\def\ea{\end{eqnarray}}

\begin{document}

\title{Vortex phase diagram of kagome superconductor CsV$_3$Sb$_5$}

\author{Xinyang Zhang}
\affiliation{Stanford Institute for Materials and Energy Sciences, SLAC National Accelerator Laboratory, 2575 Sand Hill Road, Menlo Park, CA 94025, USA}
\affiliation{Geballe Laboratory for Advanced Materials, Stanford University, Stanford, CA 94305, USA}
\affiliation{Department of Applied Physics, Stanford University, Stanford, CA 94305, USA}

\author{Mark Zic}
\affiliation{Geballe Laboratory for Advanced Materials, Stanford University, Stanford, CA 94305, USA}
\affiliation{Department of Physics, Stanford University, Stanford, CA 94305, USA}

\author{Dong Chen}
\affiliation{Max Planck Institute for Chemical Physics of Solids, 01187 Dresden, Germany}
\affiliation{College of Physics, Qingdao University, Qingdao 266071, China}

\author{Chandra Shekhar}
\affiliation{Max Planck Institute for Chemical Physics of Solids, 01187 Dresden, Germany}

\author{Claudia Felser}
\affiliation{Max Planck Institute for Chemical Physics of Solids, 01187 Dresden, Germany}

\author{Ian R. Fisher}
\affiliation{Stanford Institute for Materials and Energy Sciences, SLAC National Accelerator Laboratory, 2575 Sand Hill Road, Menlo Park, CA 94025, USA}
\affiliation{Geballe Laboratory for Advanced Materials, Stanford University, Stanford, CA 94305, USA}
\affiliation{Department of Applied Physics, Stanford University, Stanford, CA 94305, USA}

\author{Aharon Kapitulnik}
\affiliation{Stanford Institute for Materials and Energy Sciences, SLAC National Accelerator Laboratory, 2575 Sand Hill Road, Menlo Park, CA 94025, USA}
\affiliation{Geballe Laboratory for Advanced Materials, Stanford University, Stanford, CA 94305, USA}
\affiliation{Department of Applied Physics, Stanford University, Stanford, CA 94305, USA}
\affiliation{Department of Physics, Stanford University, Stanford, CA 94305, USA}

\date{\today}

\begin{abstract}
 The screening response of vortices in kagome superconductor CsV$_3$Sb$_5$ was measured using the ac mutual inductance technique. Besides confirming the absence of gapless quasiparticles in zero external magnetic field, we observe the peak effect, manifested in enhanced vortex pinning strength and critical current, in a broad intermediate range of magnetic field. The peaks are followed by another crossover from strong to weak pinning, unlike the usual peak effect that diminishes smoothly at $H_{c2}$. Hysteresis in the screening response allows the identification of a vortex glass phase which strongly correlates with the onset of the peaks. A variety of features in the temperature- and field-dependence of the screening response, corroborated by resistance and dc magnetization measurements, have allowed us to extract an $H$-$T$ phase diagram of the vortex states and to infer the irreversibility line $H_\text{irr}(T)$.
\end{abstract}

\maketitle

Correlated metals in reduced dimensions often exhibit propensity to multiple ordered phases with similar energies, which leads to the phenomenon of ``intertwined order’’ \cite{Fradkin2015,Fernandes2019}, where multiple phases emerge out of a primary phase.  Kagome metals,  featuring correlated electron effects and a topologically nontrivial band structure, have recently emerged as a compelling family of compounds to realize such multiple ordered electronic states.  A particularly interesting material system in that respect is the class of quasi-two-dimensional kagome metals \AVS, which exhibit charge order transitions at $\sim$80 K, 103 K and 94 K for A = K, Rb and Cs respectively \cite{ortiz_cs_2020}.  Focusing on \CVS, a CDW transition is revealed, which is associated with a substantial reconstruction of the Fermi surface pockets linked to the vanadium orbitals and the kagome lattice framework \cite{Ortiz2021}. Nuclear magnetic resonance studies on the different vanadium sites is consistent with orbital ordering at T$\sim$94 K induced by a first order structural transition, accompanied by electronic charge density wave (CDW) that appears to grow gradually below $T_{CDW}$, with possible intermediate subtle stacking transitions perpendicular to the kagome planes \cite{Song2022}. 

The appearance of superconductivity with $T_c$ ranging from $\sim$2.5 K to 4 K prompted further examinations of the relation between the normal and the superconducting states \cite{xu_multiband_2021, gupta_microscopic_2022}, with initial studies suggesting a pair-density-wave (PDW) \cite{Chen2021}. While the possibility of a time-reversal symmetry breaking CDW state \cite{Kun2022} is still under debate \cite{Saykin2023}, the emerging superconducting state appears to exhibit a gapped conventional s-wave order parameter \cite{Kun2022}. With in-plane $H_{c2}$ clearly dominated by orbital effects and a large Ginzburg-Landau parameter (here we find $\kappa\sim35$), we might expect a rather ordinary vortex physics behavior. Considering the non-trivial topological band structure of \CVS,  Majorana zero modes (MZM) excitations in vortex cores \cite{Fu2008} may be expected. However, it is not yet clear whether there exists a range of magnetic fields where vortex-induced MZM do not fade-out due to overlap and thus affect the observed $H$-$T$ phase diagram of this material.

In this letter we examine the vortex phase diagram of \CVS~through measurements of the superfluid response using a high-resolution mutual-inductance probe, supplemented by resistivity and magnetization measurements.  Whereas the low-temperature zero-field data confirms the absence of gapless quasiparticles, we find an unusually wide temperature-range of phase fluctuations, which at a finite magnetic field evolves into a wide vortex liquid phase.  For a constant magnetic field, an irreversibility line is observed, separating unpinned dissipative vortex liquid from a weak pinning regime characterized by weak sensitivity to the measurement time scale. The vortex lattice melting transition, appearing at lower fields and temperatures, is identified as an abrupt increase in inductive response. Finally, a pronounced ``peak effect'' is observed below the melting transition, exhibiting a hysteresis that weakens with increasing temperature. We therefore observe that \CVS provides a unique example of a clean anisotropic superconductor where a hierarchical succession of transitions and crossovers in the vortex state do not cross each other (by contrast to e.g. Bi$_2$Sr$_2$CaCu$_2$O$_{8+y}$ \cite{majer_separation_1995}) and where the melting point is clearly distinct from other effects (by contrast to e.g. YBa$_2$Cu$_3$O$_{7-y}$ \cite{Pal2000}). Taking into account the strong anisotropy of \CVS~\cite{ortiz_cs_2020,ni_anisotropic_2021}, we calculate a low temperature limit for vortex-entropy change as $\sim0.04\ k_B$ per layer. 

Peak effect is often observed in the inductive part of superfluid response, while exhibiting an enhancement in the depinning critical current density $j_c$ near $H_{c2}$ \cite{pippard_possible_1969}. Typically accompanied by a hysteresis, it is closely linked to the strengthening of pinning in a vortex glass regime \cite{larkin_pinning_1979, giamarchi_phase_1997}. Such effects have been under scrutiny in a broad range of type-II superconductors including Nb(O, Ti) \cite{de_sorbo_peak_1964}, NbZr \cite{berlincourt_superconductivity_1961}, YBa$_2$Cu$_3$O$_{7-\delta}$ \cite{ling_ac_1991, kwok_peak_1994}, 2H-NbSe$_2$ \cite{higgins_varieties_1996}, (Ba, K)Fe$_2$As$_2$ \cite{ge_peak_2013}, ternary stannide Ca$_3$Rh$_4$Sn$_{13}$ \cite{tomy_observation_1997}, as well as germanides Lu$_3$Os$_4$Ge$_{13}$ and Y$_3$Ru$_4$Ge$_{13}$ \cite{weng_nodeless_2017}. Different scenarios for the interplay between vortex pinning and lattice elasticity defines melting and irreversibility lines, which play important roles in determining the magnetic properties of high-$T_c$ cuprates such as YBa$_2$Cu$_3$O$_7$ and Bi$_2$Sr$_2$CaCu$_2$O$_8$ \cite{brandt_flux-line_1995}, often led to conflicting phase diagrams, particularly in discerning the thermodynamic melting transition out of the pinning-induced vortex state. The clear feature that we observe between the irreversibility line and the peak effect is unique in that respect, especially as it seems to mark the melting transition.

\begin{figure}[t]
	\centering
	\begin{subfigure}{}
		\includegraphics[width=0.48\columnwidth]{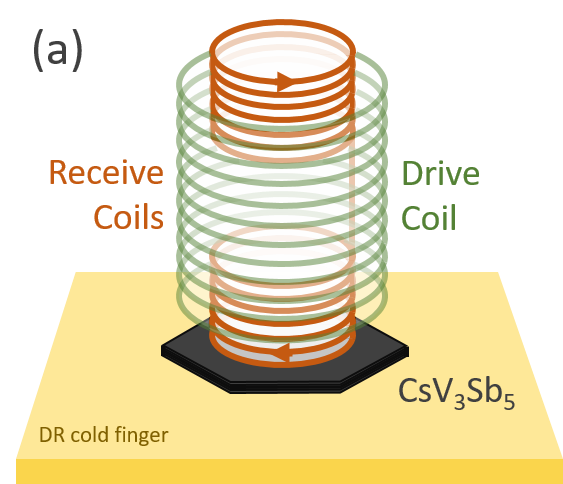}
		\label{fig1-1}
	\end{subfigure}
	\begin{subfigure}{}
		\includegraphics[width=0.47\columnwidth]{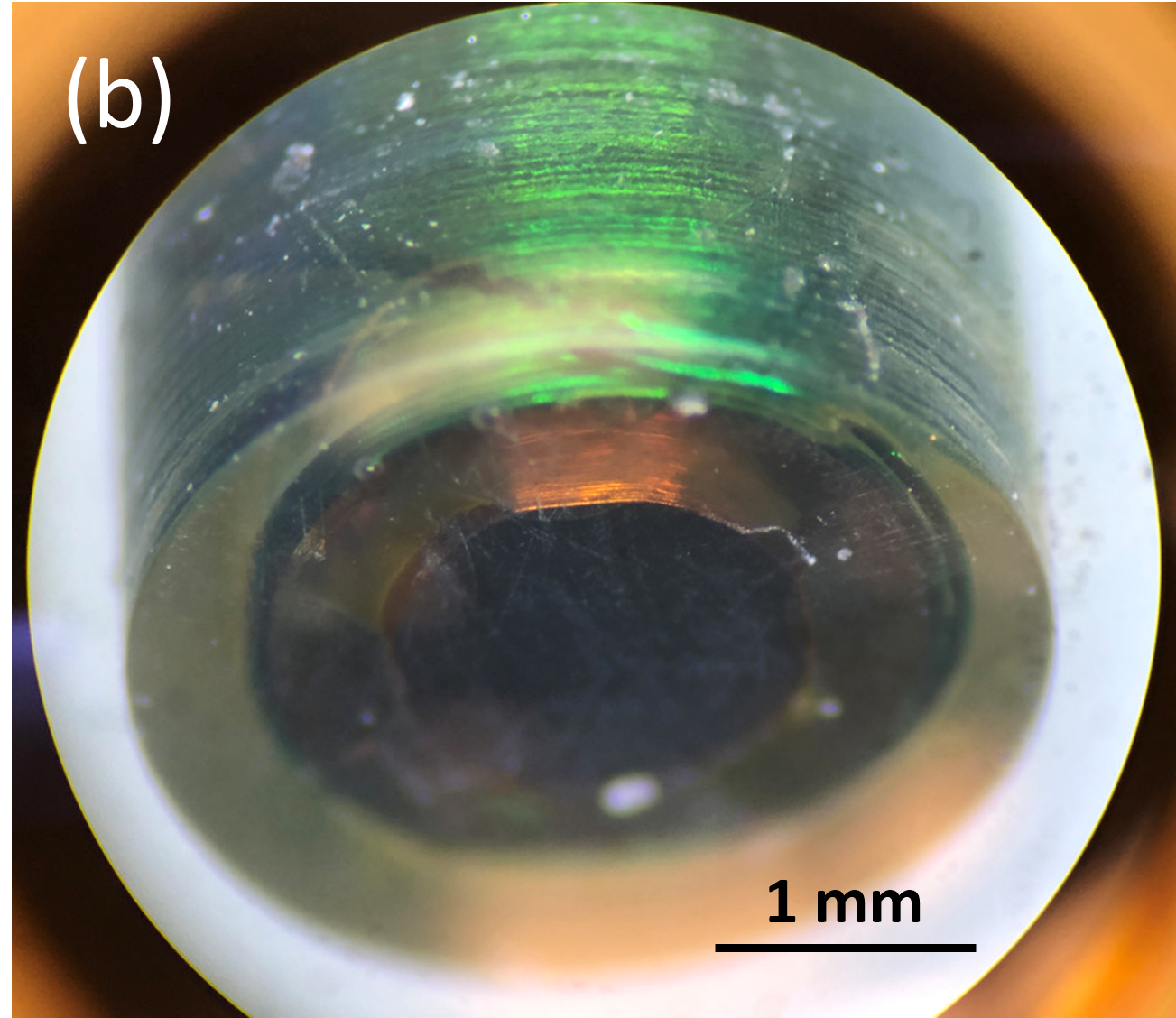}
		\label{fig1-2}
	\end{subfigure}
	\caption{The mutual inductance (MI) measurement. (a) Simplified 3d view of the MI probe mounted over a CsV$_3$Sb$_5$ sample, which is thermally anchored by a thick puddle of silver paint onto the mixing chamber of a dilution refrigerator. (b) Bottom view of the MI probe under the microscope. The green-colored drive coil and bronze-colored receive coils were cast in epoxy, which is polished to $\sim 20\ \mu$m between the wires and the bottom surface. The scale bar indicates 1 mm.}
	\label{fig1}
\end{figure}

The mutual inductance (MI) technique that we are adopting for our studies offers an alternative approach of measuring the full complex ac superfluid response \cite{fiory_penetration_1988, jeanneret_inductive_1989}. In zero external magnetic field, screening response due to a small ac drive can be related to superfluid density and (effective) penetration depth, given exact geometries of the sample and MI coils \cite{turneaure_numerical_1996}. In the present study, the complex ac response $V=V'-iV''$ of superconducting \CVS~ single crystals was measured using a gradiometer-type MI probe \cite{jeanneret_inductive_1989} as shown in Fig.~\ref{fig1}. In this configuration, both the drive coil and the astatically-wound pair of receive coils (1.5-mm in diameter) are positioned above the sample. The drive coil is supplied an ac current of $20\ \mu$A at $100$ kHz, thus creating a minuscule ac magnetic field $H_{ac}\lesssim3\ \text{mOe}\ll H_{c1}$ and inducing screening supercurrents (and potentially dissipations) in the superconductor; meanwhile, the receive coils pick up both quadratures $V'$ and $V''$ of the induced emf using a lock-in amplifier phase-locked at the drive frequency.

CsV$_3$Sb$_5$ single crystals were prepared by self-flux method \cite{ortiz_cs_2020,Chen2022}. An approximately $2\times2\times0.1$ mm$^3$ hexagonal-shaped CsV$_3$Sb$_5$ crystal was silver pasted onto the cold finger of a dilution refrigerator, where the MI coils are mounted above and gently pressed down onto the sample. Care was taken to minimize the stress exerted by the coil assembly, which together with the measurement wiring is well-thermalized at the mixing chamber temperature. Details of the coil assembly can be found in the supplemental information.

\begin{figure}[t]
	\centering
	\includegraphics[width=\columnwidth]{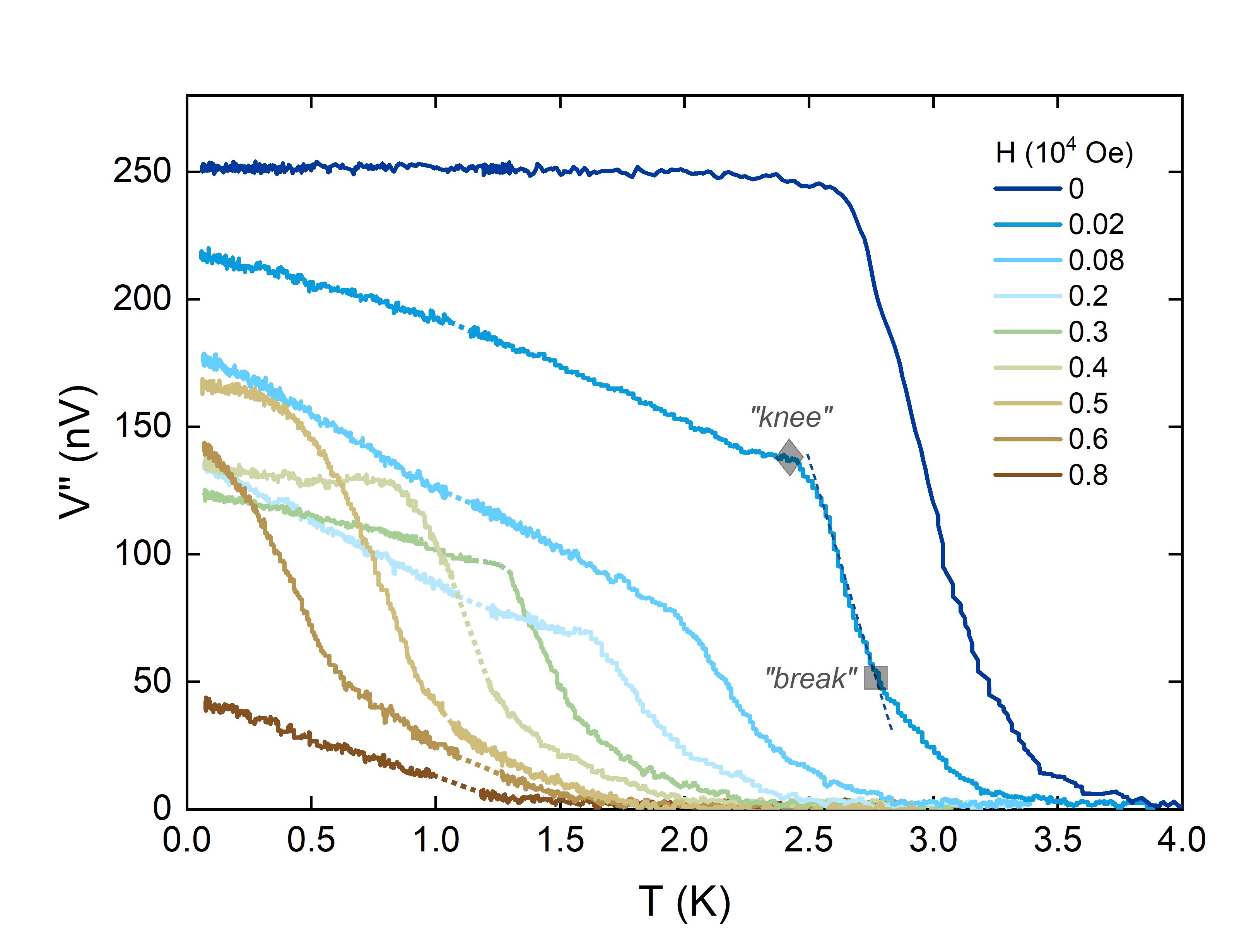}
	\caption{Temperature dependence of inductive response $V''(T)$ in varying perpendicular magnetic field. Dotted lines connect data $< 1$ K and $> 1.2$ K. A ``knee'' feature is apparent in almost every curve and its location is non-monotonic as field increases. Above the ``knee,'' a rapid linear drop in signal crosses over to a more gradual decrease to zero, separated by a ``break'' in behavior. For further explanations see Fig.~\ref{fig4}.}
	\label{fig2}
\end{figure}

We first focus on the superconducting transition in zero magnetic field as shown in Fig.~\ref{fig2}. With careful tuning of the applied magnetic field such that any effect of remnant field is eliminated, we identify a zero-field superconducting transition temperature $T_c(H=0)\approx 4.0$ K, where the onset of inductive response $V''$ at 100 kHz is located. Following a gradual increase of $V''(T)$ over a transition region of $\Delta T/T_c\sim 1/4$, the inductive signal saturates at the maximum value below $T\lesssim 2$ K. The finite size of the sample with respect to our coils limited our ability to extract an effective penetration depth $\lambda_{\text{eff}}(T)\propto T^{-0.5}\exp(-\Delta(0)/T)$ or a superfluid density $n_s(T)\propto\lambda_{\text{eff}}^{-2}(T)$ from the measured $V''(T)$. Nevertheless, such temperature independence as $T\to 0$ is supported by the recent measurements of penetration depth that yielded similar results, \textit{i.e.} nodeless superconductivity exhibiting no signature of gapless quasiparticles in zero magnetic field \cite{duan_nodeless_2021, zhang_nodeless_2023}. 

Upon applying a dc magnetic field along the c-axis $H_\text{dc}>H_{c1}\sim 30$ Oe at $0.12$ K, the inductive response $V''(T)$ is increasingly suppressed and becomes temperature-dependent as $T\to 0$ in Fig.~\ref{fig2}. From the measured $H_{c1}$ and an estimated $H_{c2,\parallel c}$ \cite{ni_anisotropic_2021}, we find an effective penetration depth $\lambda_{\text{eff}}=\sqrt{\Phi_0\ln(\lambda/\xi)/4\pi}\approx 400$ nm, consistent with the tunneling diode oscillator (TDO) method \cite{duan_nodeless_2021}. In intermediate fields $2\ \text{kOe}\lesssim H\lesssim 6$ kOe, the evolution of $V''(T)$ appears non-monotonic as a function of magnetic field. For each curve below $6$ kOe, there also appears to be a ``knee'' marking a change in slope and a ``break'' that separates a gradual superconducting transition at high temperatures and a more rapid linear form at intermediate temperatures. Above $6$ kOe, $V''(T)$ drops monotonically to zero as the inductive superfluid response is increasingly suppressed. We will examine such a non-monotonic evolution in details by exploring the magnetic-field-dependence of $V''$ isotherms. 

\begin{figure}[t]
	\centering
	\includegraphics[width=\columnwidth]{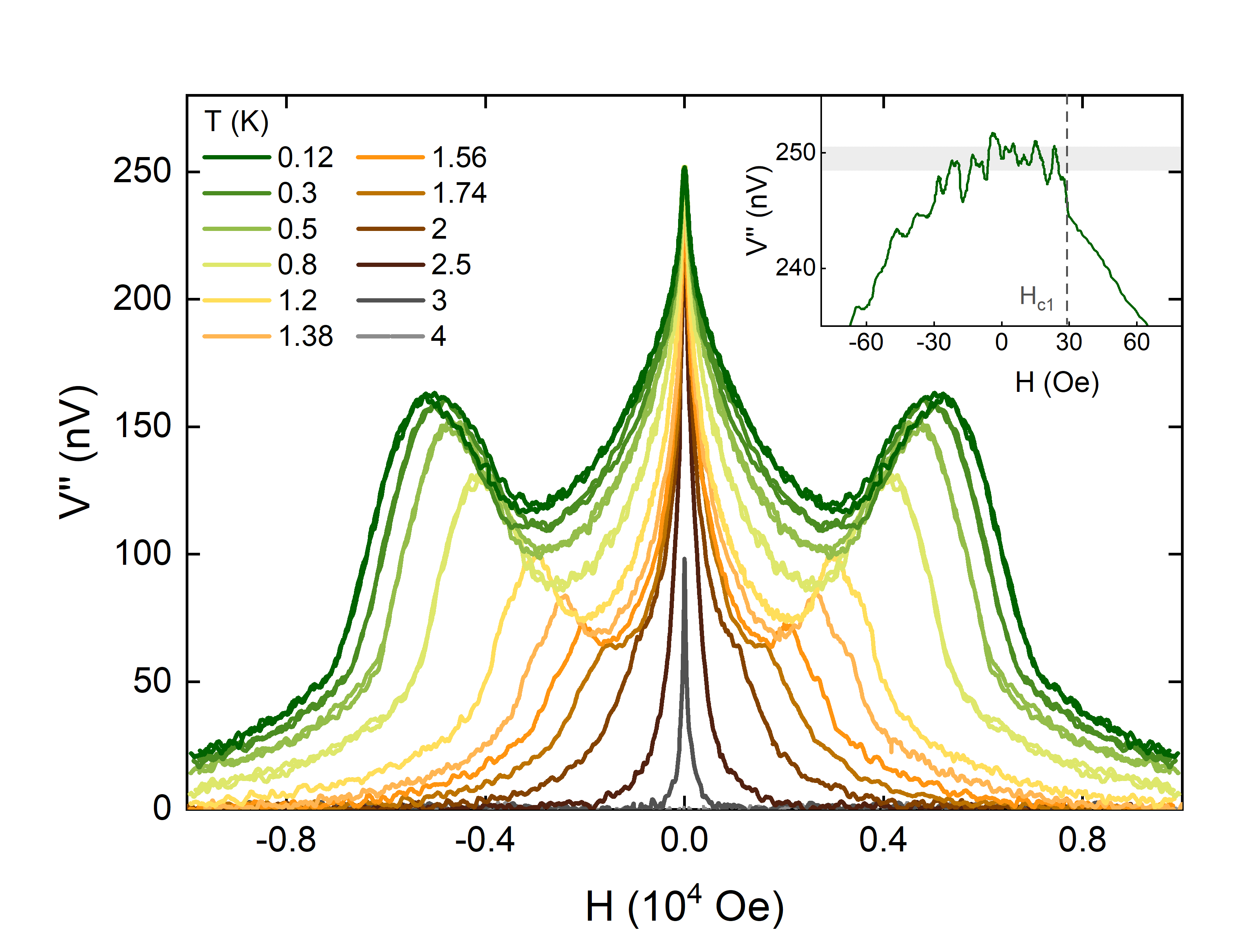}
	\caption{Magnetic field dependence of inductive response isotherms $V''(H)$ at different temperatures. External magnetic field is along the c-axis perpendicular to the V plane. For temperature below 0.5 K, both up-sweep and down-sweeps are plotted. The only observable hysteresis is at the low-field edges of the peaks, which altogether forms an envelop shape. Note the highly-symmetric shape in positive and negative field. (Inset) Expanded view of $V''(H)$ in low fields showing field-independent screening response as a signature of the Meissner state. The shadow indicates experimental uncertainty $\sim2$ nV.}
	\label{fig3}
\end{figure}

In Fig.~\ref{fig3}, isothermal field sweeps in CsV$_3$Sb$_5$ demonstrate a broad array of qualitative features related to various vortex states. Below a lower critical field $H_{c1}\sim 30$ Oe at low temperatures, the inductive ac response $V''(H)$ is independent of magnetic field as in the Meissner state. Above $H_{c1}$, $V''(H)$ drops sharply, corresponding to a rapidly weakening screening response, until a peak develops in each of the inductive response isotherms below $\sim$2 K. These peaks span a wide portion of the intermediate field range and correspond to the non-monotonic behavior in the temperature-dependence, reminiscent of the widely-known peak effect \cite{berlincourt_superconductivity_1961, de_sorbo_peak_1964, pippard_possible_1969, brandt_flux-line_1995, higgins_varieties_1996, tomy_observation_1997, ge_peak_2013, weng_nodeless_2017}.

The peaks in CsV$_3$Sb$_5$ appear more prominent at lower temperatures, approaching $30$\% of the maximal zero-field signal, while a peak was also found in critical current measurements (see supplementary). The peak's onset field ($H_{\text{onset}}$), peak field ($H_{\text{peak}}$), and width ($|H_{\text{base}}-H_{\text{onset}}|$) all increase with lower temperature, consistent with the canonical peak effect behaviors. Nevertheless, the peaks in CsV$_3$Sb$_5$ are located at an intermediate range of magnetic field, rather than close to the onset of screening response at higher fields.

Hysteresis in $V''(H)$ is largely seen at the low-field edges of the peaks with the largest split of  $\sim$7 nV between upward and downward sweeps, and can be detected until at least $0.8$ K where the split drops below measurement sensitivity. There also exists a much weaker hysteresis at lower fields (see supplemental information.) Such a hysteresis has been identified as a signature of vortex glass phase \cite{muller_flux_1987, giamarchi_phase_1997}. In addition, the low-field edges of the peaks in all isotherms together form an envelop, corresponding to a weak or an absence of temperature-dependence in the field windows where the isotherms overlap. These windows are located accordingly near the ``knees'' ($T_{\text{knee}}$) in the finite-field superconducting transition in Fig.~\ref{fig2}.

\begin{figure}[h]
	\centering
	\includegraphics[width=\columnwidth]{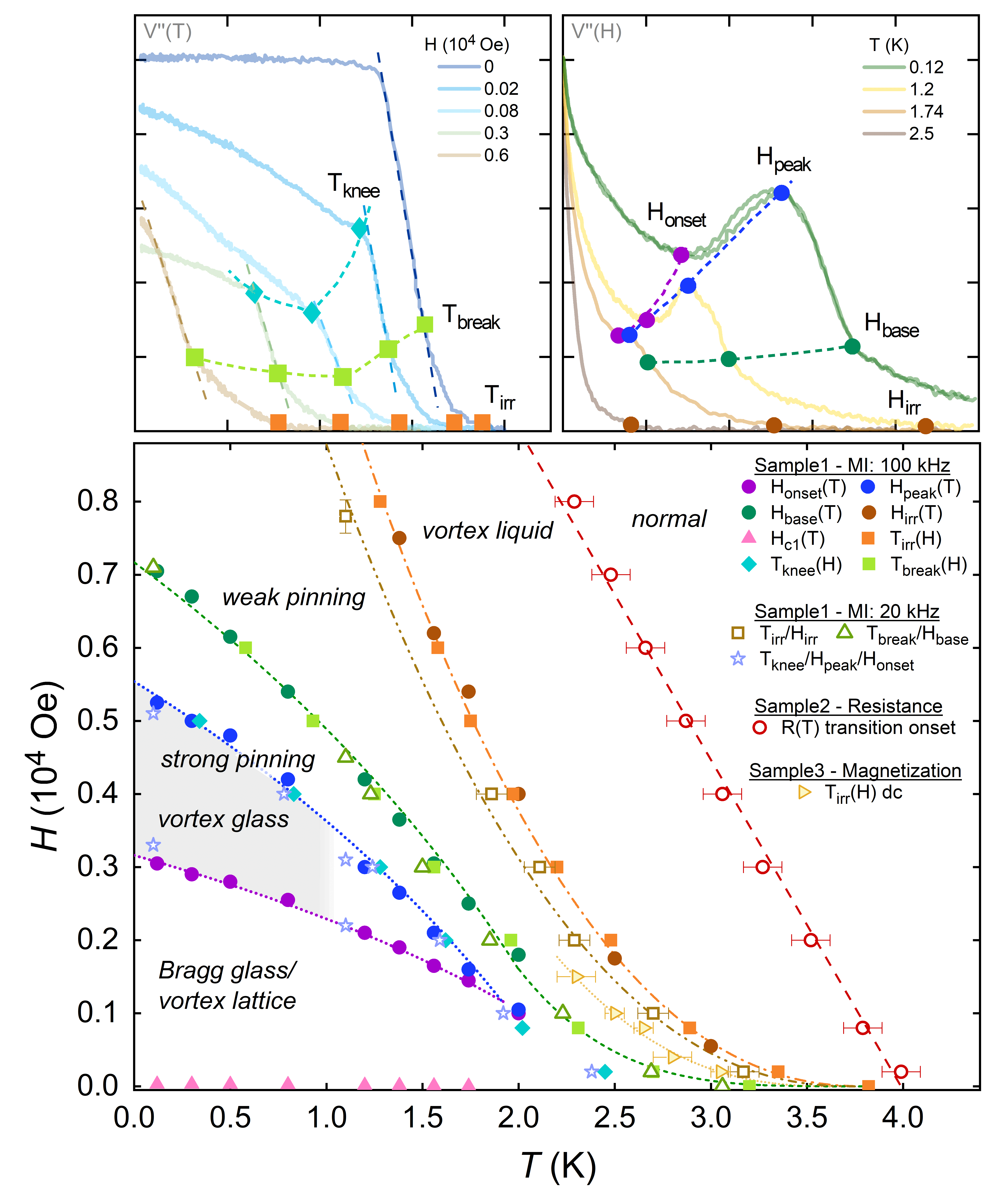}
	\caption{Vortex phase diagram of CsV$_3$Sb$_5$. (Top-left) Temperature-dependence of $V''$ at select fields showing the extraction of $T_\text{knee}$, $T_\text{break}$, and $T_c$ as explained in the main text. (Top-right) Magnetic-field-dependence of $V''$ at select temperatures showing the extraction of $H_\text{onset}$, $H_\text{peak}$, $H_\text{base}$, and $H_\text{irr}$ as explained in the main text. (Bottom) $H$-$T$ phase diagram constructed by the extracted features. All open-symbol points are extracted from 20-kHz MI measurements. The shadow marks the region with strong hysteresis. The mean-field $T_{c0}$ (red circle) is extracted from the onset of superconducting transition in resistance measurements (see supplementary). Feature extractions have uncertainties of order $\sim$0.01 K in temperature, or $\sim$100 Oe (100-kHz)/1000 Oe (20-kHz) in field (error bars suppressed for clarity except for $H_\text{irr}$ and $T_{c0}$). Extraction of $T_\text{irr}$ from dc magnetization has uncertainties $\sim 0.05-0.1$ K. Dashed lines are merely guides to the eye.}
	\label{fig4}
\end{figure}

The rich features in the temperature- and field-dependence of superfluid screening response $V''$ allow us to identify various vortex phases in an $H$-$T$ phase diagram Fig.~\ref{fig4}. Close to zero magnetic field, the Meissner state with a constant diamagnetic susceptibility is bounded by the low $H_{c1}\sim 30$ Oe and is barely visible. From the field-dependence $V''(H)$, we extract the peak's onset $H_{\text{onset}}(T)$ (violet), peak $H_{\text{peak}}(T)$ (blue), and base $H_{\text{base}}(T)$ (green) as a function of temperature. Independently, from the temperature-dependence $V''(T)$ we extract $T_{\text{knee}}(H)$ (light blue) and $T_{\text{break}}(H)$ (light green) as a function of field as explained previously. The resulting curves are surprisingly consistent with each other, where $T_{\text{knee}}(H)$ and $H_{\text{peak}}(T)$ as well as $T_{\text{break}}(H)$ and $H_{\text{base}}(T)$ are perfectly aligned, respectively. The peaks become indistinguishable above $\sim 0.5 T_c\approx 2$ K.

We begin with the low-field and low-temperature regimes of the $H$-$T$ phase diagram. The strong hysteretic behavior (shadow) is bounded between $H_{\text{onset}}(T)$ (violet) and $H_{\text{peak}}(T)$ (blue), with its magnitude dropping below our sensitivity above $\sim 1$ K. Consequently, $H_{\text{onset}}$ can be designated as the onset of a strongly hysteretic vortex glass phase, replacing a less-disordered dislocation-free ``Bragg glass'' phase at lower fields \cite{giamarchi_phase_1997}. The vortex glass region is highly correlated with the peaks in the screening response $V''(H)$ which signifies an enhancement in flux pinning \cite{tomy_observation_1997}. Such an enhancement is commonly achieved through a crossover from the Larkin-Ovchinnikov collective pinning to individual pinning of the flux lines, following the softening of vortex lattice elastic moduli \cite{larkin_pinning_1979} in the presence of a significant anisotropy \cite{ni_anisotropic_2021}. 

In previous studies of the vortex phase diagram in layered materials, particularly the cuprates, a gradual evolution in pinning strength is commonly expected between the boundary of the peak effect at $H_{\text{peak}}$ and $H_{\text{irr}}$. In these cases and with no otherwise observable features, $H_{\text{peak}}$ is identified with the melting transition (see e.g. \cite{Higgins1996,Kupfer2000}.)  However, where a magnetization discontinuity is observed without a peak effect, a melting line was found to cross the irreversibility line in Bi$_2$Sr$_2$CaCu$_2$O$_8$ \cite{majer_separation_1995}. In the present case of \CVS, we find a pronounced break in the response of the vortex system between $H_{\text{peak}}$ and $H_{\text{irr}}$, which we denote as $H_{\text{base}}(T)$. This feature is also visible in the temperature scans of the inductive response, which we identify as  $T_{\text{break}}(H)$. Such a break was previously found in surface impedance measurements of Bi$_2$Sr$_2$CaCu$_2$O$_8$ \cite{Hanaguri1999} and was identified with the melting line. Similar to our case, their technique, while performed at higher frequency, was sensitive to the superfluid density, which is expected to exhibit a sharp increase in superfluid density as the field is lowered through the melting transition. It is therefore natural  to  identify the line of ($T_{\text{break}}(H)$, $H_{\text{base}}(T)$) as the melting transition \cite{blatter_vortices_1994}.  This now allow us to calculate the entropy change through the transition. Following Zeldov {\it et al.}, \cite{Zeldov1995}we can use Clausius-Clapeyron equation to calculate the latent heat of the transition: $L=T_m\Delta S=-(\Delta B/4\pi)(dH_m/dT)T_m$, where $(T_m,H_m)$ is the melting line and $\Delta B$ is the discontinuity in magnetic induction at the first order melting transition. Using a standard Lindeman criterion for melting \cite{blatter_vortices_1994} to estimate $\Delta B$, we find \cite{Zeldov1995} $\Delta S\approx 0.1 d\gamma\sqrt{B_m/\Phi_0}$, where $\gamma\approx \sqrt{600}$ is the anisotropy of \CVS~and $d\approx 9.3$ \AA~is the interlayer distance \cite{ortiz_cs_2020}. At low temperatures and high fields we find that $\Delta S\approx 0.04\ k_B$ per kagome layer, which is a factor of 10 smaller than the expected value from fully decoupled layers  \cite{Hetzel1992}. This implies correlations along the vortex line, which thus limit the number of degrees of freedom per vortex.  

Returning to the peak effect, when compared with typical characteristics in both layered and non-layered superconductors \cite{de_sorbo_peak_1964,berlincourt_superconductivity_1961,ling_ac_1991, kwok_peak_1994, higgins_varieties_1996, ge_peak_2013, tomy_observation_1997, weng_nodeless_2017}, the peak effect in CsV$_3$Sb$_5$ appears more prominent and spans a large range in parameter space. The details of pinning mechanisms are still unclear in CsV$_3$Sb$_5$, whether being randomly distributed point-like impurities, or correlated effects related to the underlying CDW and its associated 
distortions. However, at least down to $\sim T_c/2$ the temperature dependence of the irreversibility line and peak effect line follow the melting line, which could indicate that in that regime surface barrier effects are not important and bulk pinning dominate \cite{majer_separation_1995}.

 In the high-field and high-temperature regimes, we attribute the onset of $V''$ ($H_\text{irr}(T)$), for different frequencies, including dc, to the presence of pinned vortices, and thus an ac ``irreversibility line'' or ``depinning line'' $H_{\text{irr}}(T)\sim (T_c-T)^\alpha$, where the exponent typically ranges from $\frac{4}{3}$ to 2 \cite{brandt_flux-line_1995}. As we are measuring the response of a coherent sum of screening currents over a select range of wave vectors determined by the receive coil geometry \cite{jeanneret_inductive_1989} and over a time scale determined by the drive frequency, our measurement is mostly sensitive to flux lines that appear pinned within the experiment's parameters. In that respect, the region bounded by the irreversibility and peak effect lines indicate the increase in pinning efficiency, starting from thermally activated flux flow (TAFF) \cite{malozemoff_frequency_1988} at high temperatures, with its hallmark of frequency dependence (see Fig.~\ref{fig4}) to the strong pinning in the peak effect regime with intermediate change in pinning behavior below the freezing of the vortex lattice.  
 
 Finally we note that debates persist for an unambiguous identification of the irreversibility line since it is highly sensitive to either geometrical barriers or the external drive current that causes depinning of vortices \cite{majer_separation_1995,brandt_flux-line_1995}. It is important to note that the drive field in our experiment is several orders of magnitude lower compared to typical ac susceptibility measurements at $\sim 1$ Oe, whereas our MI experiment is complementary to transport measurements that typically require much higher drive current. Further clarifications on the irreversibility line may be achieved through torque magnetometry \cite{farrell_magnetization_1996} or local magnetic induction experiments\cite{majer_separation_1995}.

In summary, the screening response of vortices in kagome superconductor CsV$_3$Sb$_5$ has been measured using the ac mutual inductance technique. Besides confirming the absence of gapless quasiparticles in zero external magnetic field, we observe the peak effect, corresponding to enhanced pinning strength and critical current, in a broad intermediate range of magnetic field. The peaks vanishes at a melting transition from strong to weak pinning, unlike the usual peak effect that ends near $H_\text{irr}$ or $H_{c2}$. Hysteresis in $V''$ leads us to the identify a vortex glass phase, where its onset is highly correlated with that of the peaks. The various features in the temperature- and field-dependence of the screening response, corroborated by transport and dc magnetization measurements, have allowed us to construct an $H$-$T$ phase diagram of the vortex states and to infer the irreversibility line $H_\text{irr}(T)$. \\

\noindent {\bf Acknowledgments:} Work at Stanford University was supported by the Department of Energy, Office of Basic Energy Sciences, under contract no. DE-AC02-76SF00515. The authors gratefully acknowledge (financial) support by the Deutsche Forschungsgemeinschaft through SFB 1143 (Project-ID 258499086) and through the Würzburg-Dresden Cluster of Excellence on Complexity and Topology in Quantum Matter ct.qmat (EXC 2147, Project-ID 39085490).

\bibliography{CVS_MI}

\onecolumngrid
\newpage
\setcounter{section}{0}
\setcounter{figure}{0}
\renewcommand{\thefigure}{S\arabic{figure}}
\renewcommand{\theequation}{S.\arabic{equation}}
\renewcommand{\thetable}{S\arabic{table}}
\renewcommand{\thesection}{S\arabic{section}}

\renewcommand{\thefootnote}{\fnsymbol{footnote}}

\begin{center}
	\textbf{SUPPLEMENTAL INFORMATION for} \\
	\vspace{1em}
	\textbf{Vortex phase diagram of kagome superconductor CsV$_3$Sb$_5$}\\
	
	\fontsize{9}{12}\selectfont
	
	\vspace{2em}
	Xinyang Zhang,$^{1,2,3}$ Mark Zic,$^{2,4}$ Dong Chen,$^{5,6}$ Chandra Shekhar,$^{5}$\\Claudia Felser,$^{5}$ Ian R. Fisher,$^{1,2,3}$ and Aharon Kapitulnik,$^{1,2,3,4}$\\
	\vspace{1em}
	$^1${\it Stanford Institute for Materials and Energy Sciences,\\ SLAC National Accelerator Laboratory, 2575 Sand Hill Road, Menlo Park, CA 94025, USA}\\
	$^2${\it Geballe Laboratory for Advanced Materials, Stanford University, Stanford, CA 94305, USA}\\
	$^3${\it Department of Applied Physics, Stanford University, Stanford, CA 94305, USA}\\
	$^4${\it Department of Physics, Stanford University, Stanford, CA 94305, USA}\\
	$^5${\it Max Planck Institute for Chemical Physics of Solids, 01187 Dresden, Germany}\\
	$^6${\it College of Physics, Qingdao University, Qingdao 266071, China}\\
\end{center}

\section{List of supplemental contents:}
\begin{enumerate}[label=\Roman*.]
	\item Details about the mutual inductance coil assembly
	\item Detailed low-field inductive response $V''(H)$ and the extraction of effective penetration depth
	\item In-phase (dissipative) component $V'$ of ac response
	\item High-quality measurements of hysteresis in field-dependent screening response
	\item Temperature- and field-dependence of $V''$ at 20 $\text{kHz}$
	\item Resistance and current-voltage characteristics of C$\text{s}$V$_3$S$\text{b}_5$
	\item Peak effect in critical current at 1.88 K
	\item Temperature-dependence of dc magnetization
\end{enumerate}

\newpage
\section{I. Details about the mutual inductance coil assembly}
\begin{figure}[h]
	\centering
	\begin{subfigure}{}
		\includegraphics[width=0.35\columnwidth,angle=270,origin=c,trim=8cm 0 5.2cm 2cm,clip]{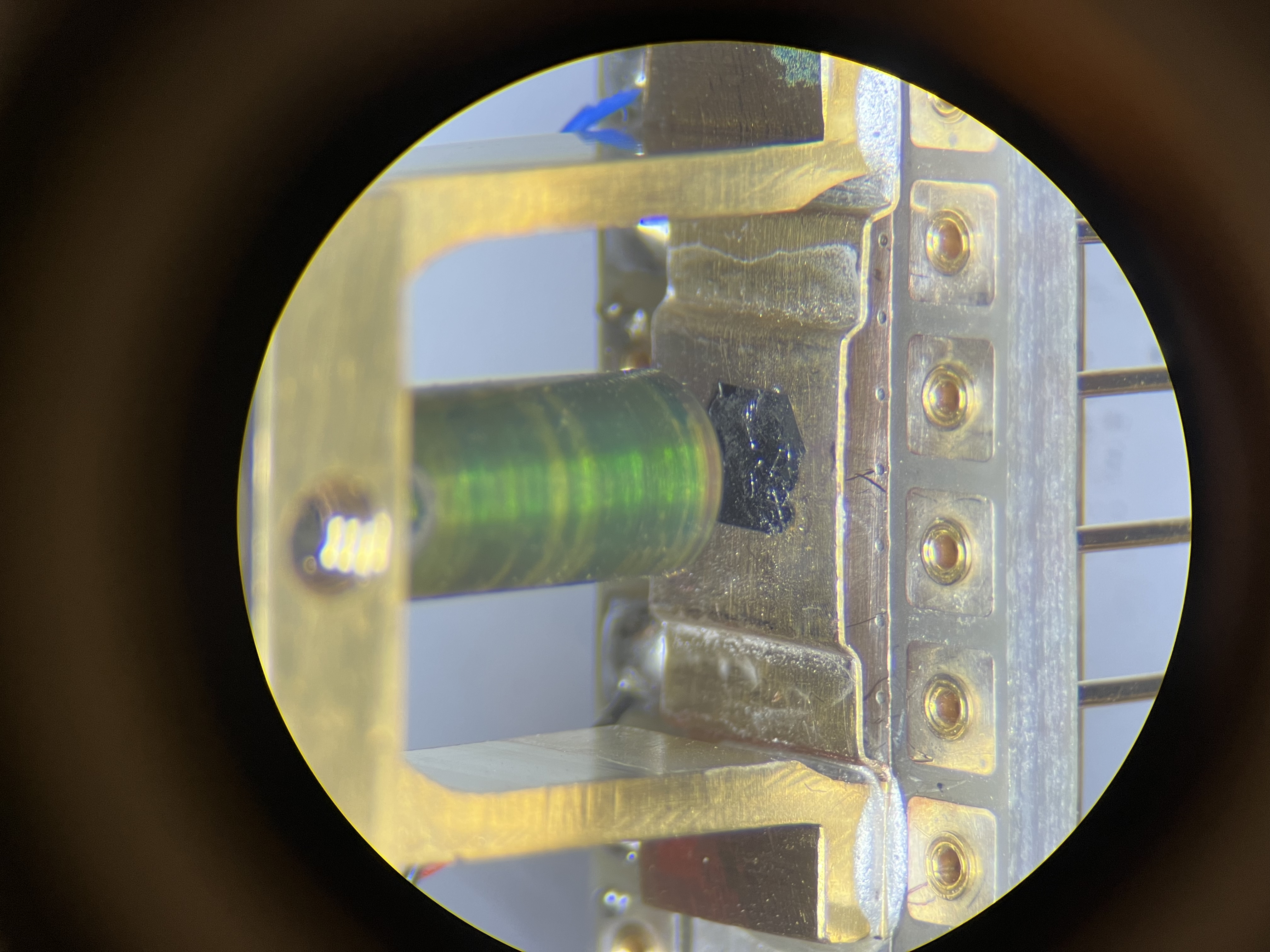}
		\label{figs1-1}
	\end{subfigure}
	\begin{subfigure}{}
		\includegraphics[width=0.5\columnwidth,origin=c,trim=1cm 1cm 2.5cm 1cm,clip]{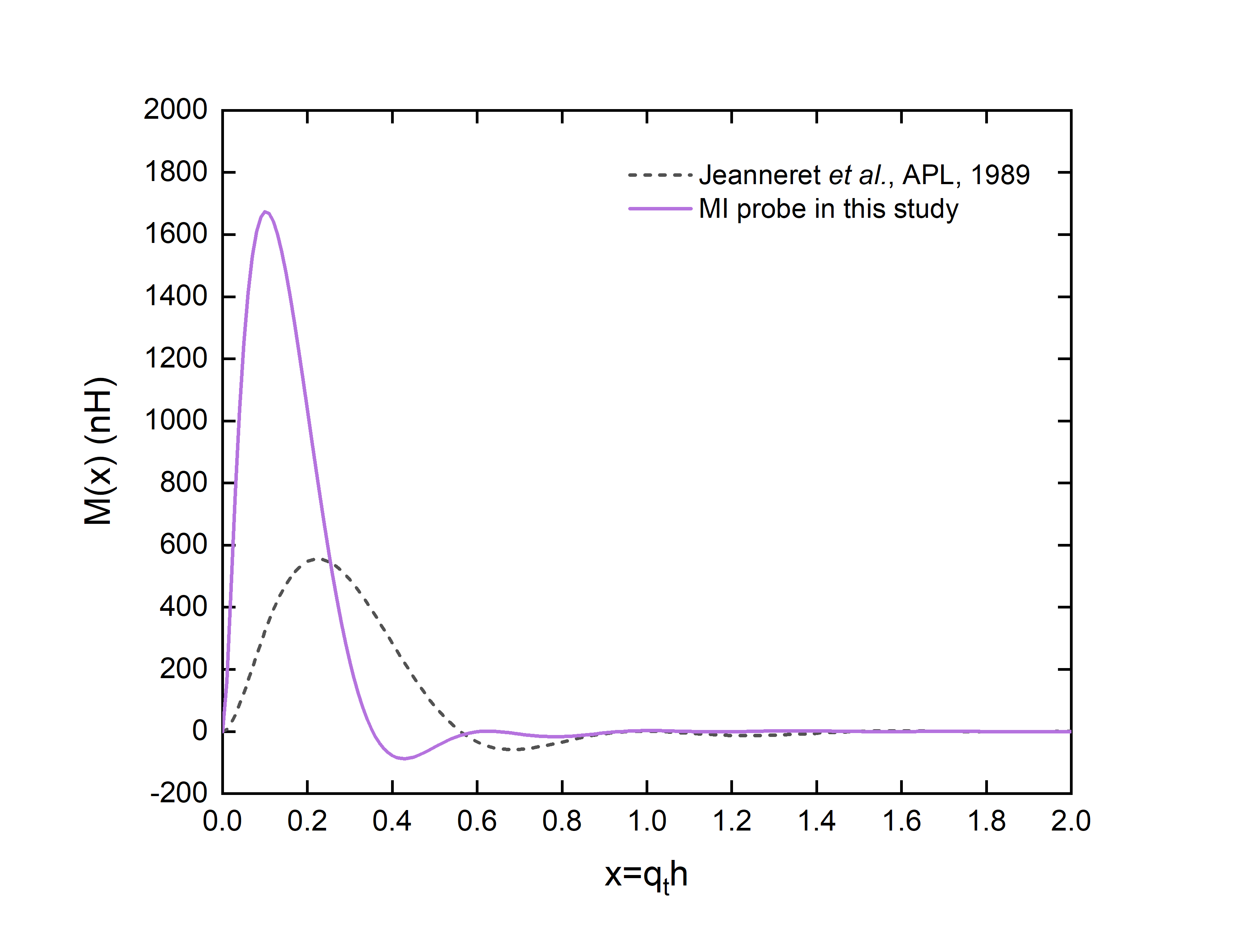}
		\label{figs1-2}
	\end{subfigure}
	\begin{subfigure}{}
		\includegraphics[width=\columnwidth]{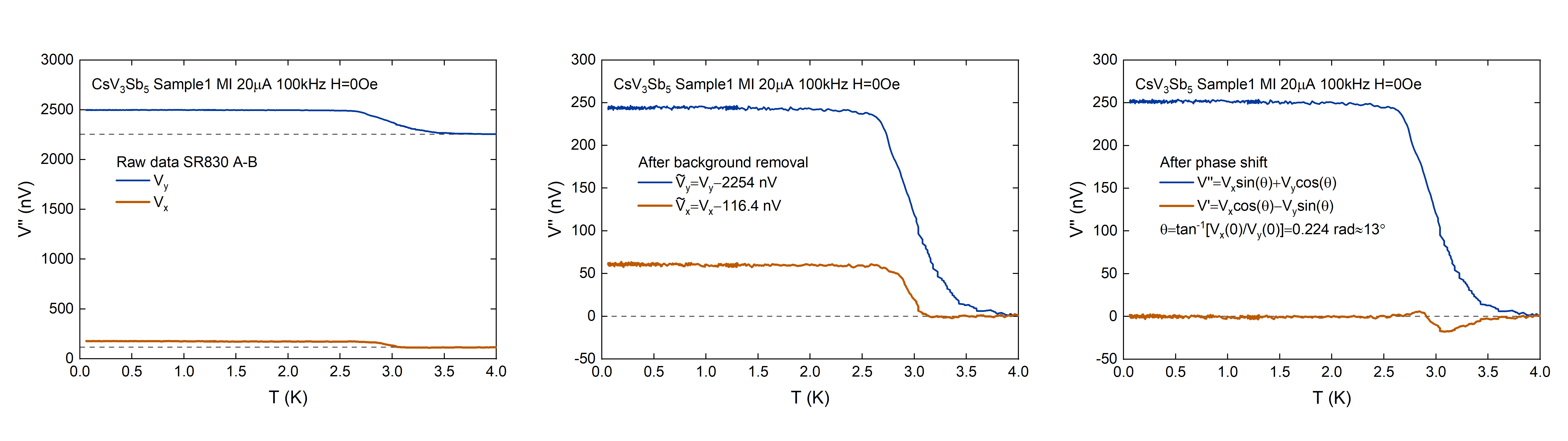}
		\label{figs1-3}
	\end{subfigure}
	\label{figs1}
	\caption{MI experiment and data analysis. (Upper left) MI probe used in this study under the microscope. The entire coil assembly was cast in epoxy and is mounted on a gold-plated copper mount. The platform where the sample is mounted on is a transfer-molded gold-plated copper spring thermally anchored to a chip carrier. (Upper right) The mutual inductance function $M(x)$ as a function of dimensionless transverse spatial wave vector, explained in the supplementary text. Dashed line is $M(x)$ for the pioneering coil used in Jeanneret \textit{et al.} (Lower left) MI data processing procedure with an example data set in zero field at 100 kHz. The raw data are shown as directly measured in both quadratures of an SR830 lock-in amplifier. (Lower center) After removing a constant background, the two quadratures were set to zero $>4$ K. (Lower right) After a phase shift, the signal was decomposed into real and imaginary parts based on the assumption that $V'(H=0, T=0)=0$.}
\end{figure}
Our mutual inductance (MI) probe is of the gradiometer-type consisting of compensated receive coils and a drive coil positioned co-axially with each other. The drive coil is supplied with an ac current of 20 $\mu$A, via a 100 k$\Omega$ current-limiting resistor. The resulting ac magnetic field is screened by the superconductor, which can be characterized by a complex ac conductance, producing a screening current in response. The magnetic field associated with this screening current in turn induces an emf in the receive coils, whose potential drop is measured by a lock-in amplifier. The induced voltage for a certain coil geometry (captured within the mutual inductance function $M(x)$) is given by: (assuming sample as an infinite plane)
\begin{equation}
	\delta V=i\omega I_D\int_0^\infty dx\ \frac{M(x)}{1+2x/i\mu_0 h\omega G}
	\label{S1}
\end{equation}

The geometrical parameters of our coils are: (receive) radius $r_R=0.75$ mm; wire diameter $\delta h_R=0.02$ mm; number of turns $N_R=56$; distance to sample $h_R\approx 0.02$ mm; (drive) radius $r_D=1.25$ mm; wire diameter $\delta h_D=0.08$ mm; number of turns $N_D=71$; distance to sample $h_D\approx 0.1$ mm. The coil assembly is based on a machined Nylon base, and was encapsulated in Stycast 1266 epoxy for mechanical integrity. The whole coil assembly is screwed into a gold-plated copper stand, which also serves as a good thermal anchor for the coil assembly. Electrical connections to the coils were made by silver epoxy. The sample was loaded on the cold finger into a thick layer of silver paint to reduce the stress exerted by the coil on the sample.

Fig.~S1 shows the calculated mutual inductance function, \textit{i.e.} a geometrical factor that determines the sensitivity of the MI probe.
\begin{equation}
	M(x)=\pi\mu_0 h\alpha\beta J_1(\alpha x)J_1(\beta x)e^{-x}\frac{1-e^{N_D\gamma x}}{1-e^{-\gamma x}}\frac{1-e^{N_R\delta x}}{1-e^{-\delta x}}
	\label{S2}
\end{equation}
where $h=h_R+h_D$ is the sample-coil distance, the dimensionless $x=q_t h$ is the transverse spatial wave vector in units of $1/h$, and $\alpha,\beta,\gamma,\delta$ are $R_D,R_R,\delta h_D,\delta h_R$ in units of $h$, respectively. $J_1(x)$ is the $n=1$ Bessel function of the first kind. With our coil geometry, the calculated mutual inductance function generally behaves as a Bessel function, where the inductance starts out as zero at $x=0$, i.e. insensitive to spatially uniform magnetic flux. The largest inductance occurs at $x=1/4$, where $q_t=1/4h\sim 2.5$ mm$^{-1}$, corresponding to a wavelength roughly twice the diameter of our receive coil corresponding to the maximal flux. At $x=1/2$ the wave vector $q_t\sim 5$ mm$^{-1}$, corresponding to the diameter of our receive coil, so the flux equals zero. For longer wavelengths the signal is largely canceled and therefore we are mostly sensitive to magnetic fields produced by the screening current with a wave vector around $q_t=1/4h$.

In this study, due to the finite size of the CsV$_3$Sb$_5$ crystal relative to the receive coils, Eqs.~\ref{S1} and \ref{S2} cannot be directly applied for calculating the complex conductance. However, qualitative features can still be extracted from the measured temperature- and field-dependent superfluid response. Furthermore, due to the insufficient compensation between the astatic receive coil windings and any stray capacitive coupling between the coaxial transmission lines for coil wiring, there exists a \textit{temperature-independent} background to either quadrature and a phase shift that mixes the two quadratures. In Fig.~S1 we present the data processing procedure that converts the measured voltage to in-phase and out-of-phase responses of the sample. Here we assume that the dissipative response equals zero at $T=0$ and $H=0$, from which we extract a phase shift related to the capacitive couplings. We enforce the same phase shift for all measurements under a certain frequency. For a 100 kHz and $20\ \mu$A excitation, the phase shift $\theta\approx 0.224\ $rad. 

\newpage
\section{II. Detailed low-field inductive response $V''(H)$ and the extraction of effective penetration depth}
\begin{figure}[h]
	\centering
	\begin{subfigure}{}
		\includegraphics[width=0.6\columnwidth]{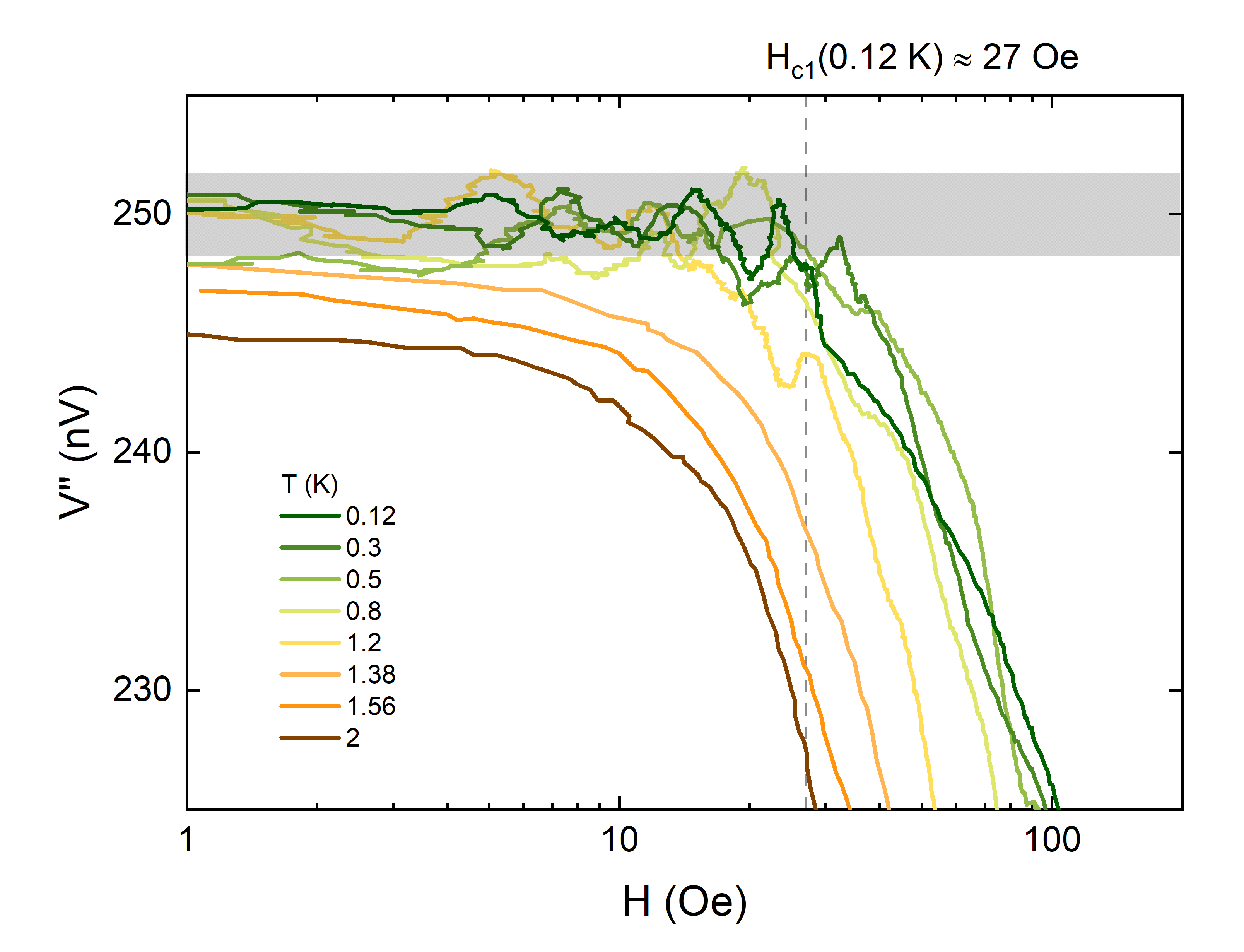}
		\label{figs2-1}
	\end{subfigure}
	\caption{In-phase (dissipative) response as a function of magnetic field (right) measured at 100 kHz. The shadow marks experimental uncertainty $\sim 2$ nV. The lower critical field at 0.12 K is identified at $\approx 27$ Oe.}
\end{figure}
The lower critical field is defined as $H_{c1}=(\Phi_0/4\pi\lambda^2)\ln(\lambda/\xi)$, where $\Phi_0=h/2e$ is the magnetic flux quantum, $\lambda$ is the penetration depth, and $\xi$ is the coherence length. In the presence of anisotropy, by taking the experimentally determined in-plane coherence length $\xi_{ab}(T=0)\approx 20$ nm, we consider the c-axis lower critical field [1]
\begin{equation}
	H_{c1\parallel c}=\frac{\Phi_0}{4\pi\lambda_{ab}^2}\ln\left(\frac{\lambda_{ab}}{\xi_{ab}}+0.5\right)\approx 30\ \text{Oe}
\end{equation}
which yields an (effective) in-plane penetration depth $\lambda_{ab}\approx 400$ nm. \\

[1] R. A. Klemm and J. R. Clem, Phys. Rev. B \textbf{21}, 1868 (1980)

\newpage
\section{III. In-phase (dissipative) component $V'$ of ac response}
\begin{figure}[h]
	\centering
	\begin{subfigure}{}
		\includegraphics[width=0.48\columnwidth]{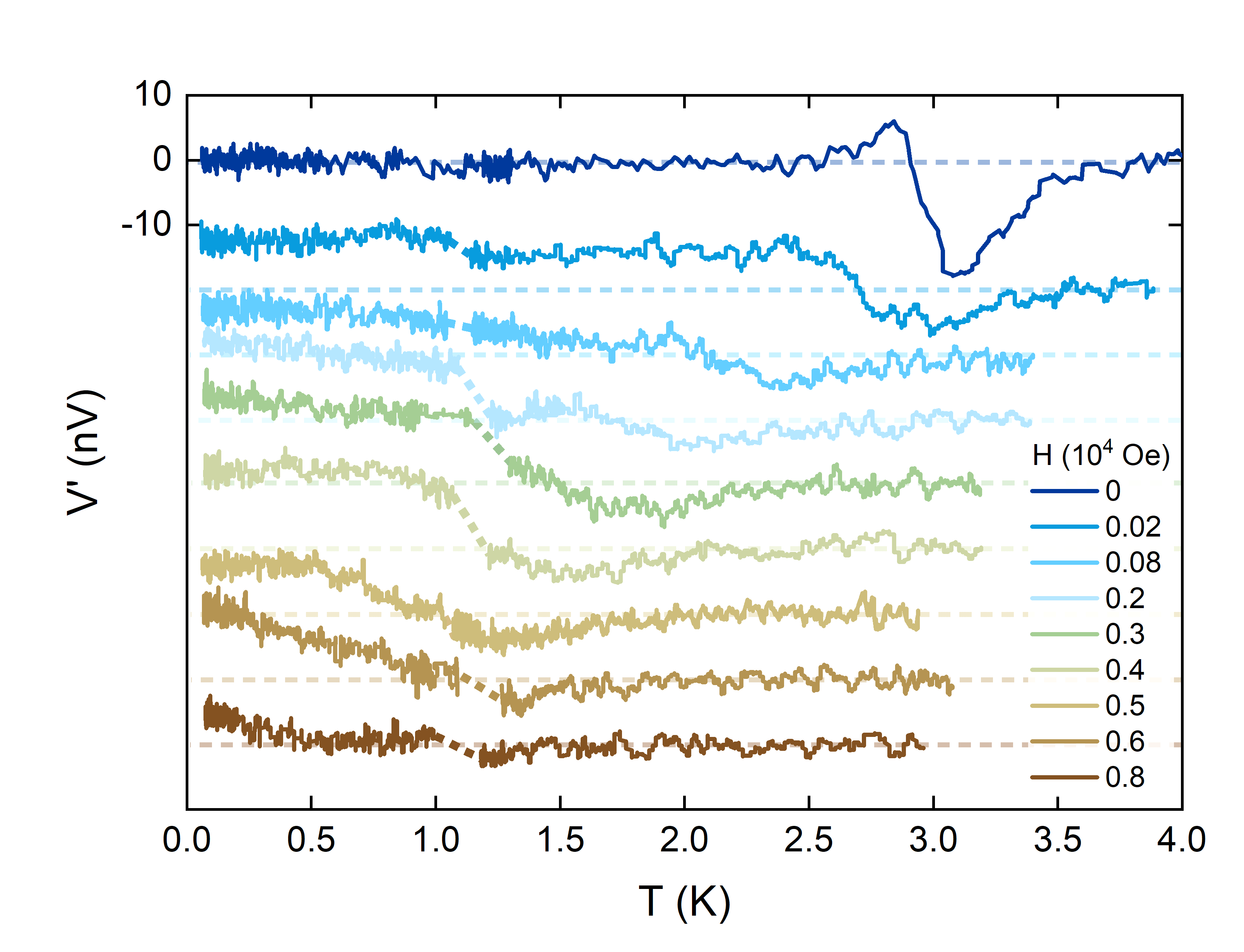}
		\label{figs3-1}
	\end{subfigure}
	\begin{subfigure}{}
		\includegraphics[width=0.48\columnwidth]{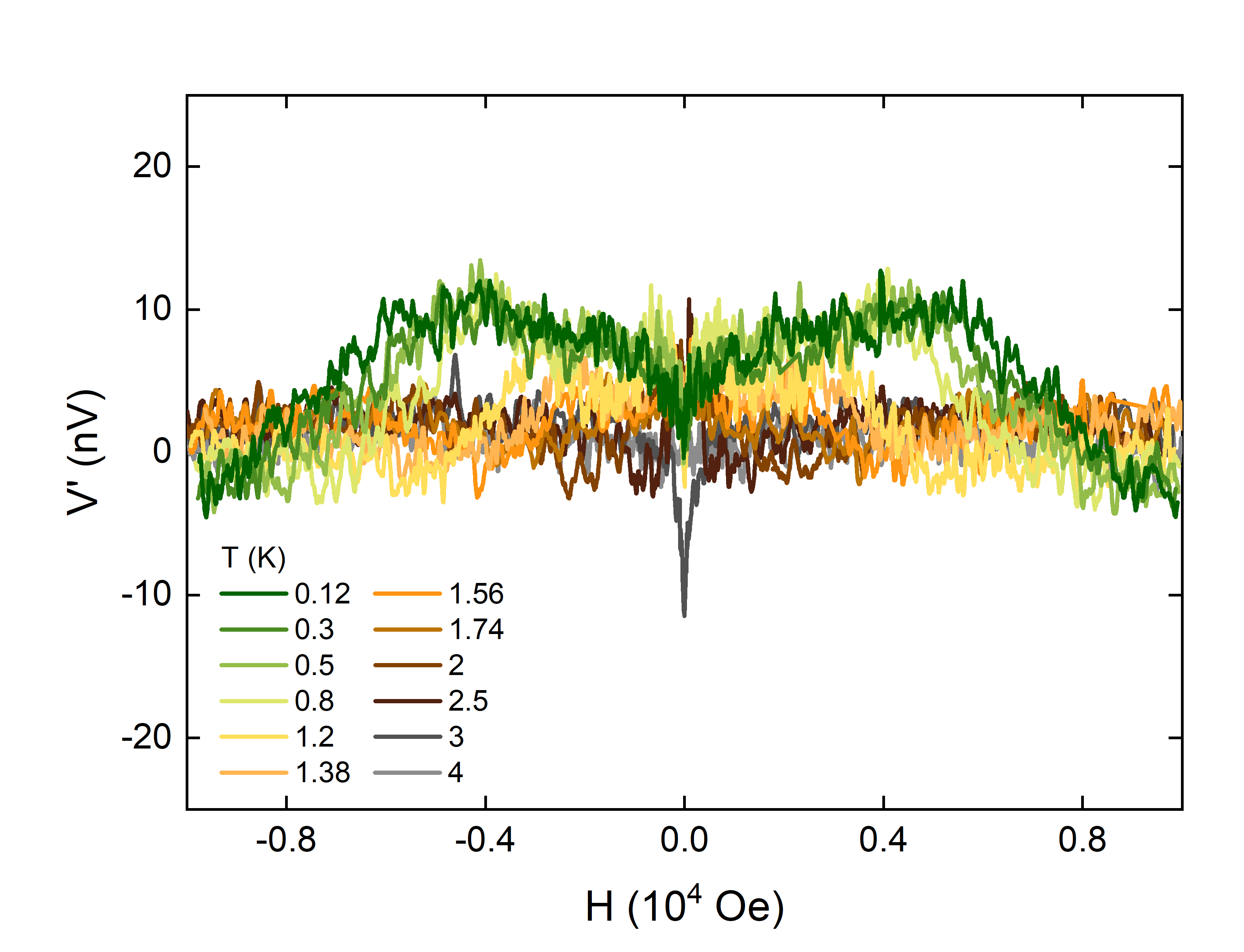}
		\label{figs3-2}
	\end{subfigure}
	\caption{In-phase (dissipative) response as a function of temperature (left) and magnetic field (right) measured at 100 kHz.}
\end{figure}
The in-phase response to the drive current is shown as a function of temperature and magnetic field. The zero level (dashed lines) of each temperature dependence has been shifted for clarity. At zero magnetic field, the superconducting is accompanied by two opposite peaks, unlike the usually loss peak that is related to the inductive response through the Kramers-Kronig relations. The dissipative response is identically zero at low temperatures, corresponding to the flattening of inductive response. This is a strong evidence that dissipation, typically associated with gapless quasi-particles, is absent at zero magnetic field. At finite field, however, the dissipative part is non-zero at low temperatures, which suggests an ac response that is lossy and not entirely inductive. The field-dependent is consistent with the temperature dependence, showing a finite dissipative response in the intermediate field range.

\newpage
\section{IV. High-quality measurements of hysteresis in field-dependent screening response}
\begin{figure}[h]
	\centering
	\begin{subfigure}{}
		\includegraphics[width=0.48\columnwidth]{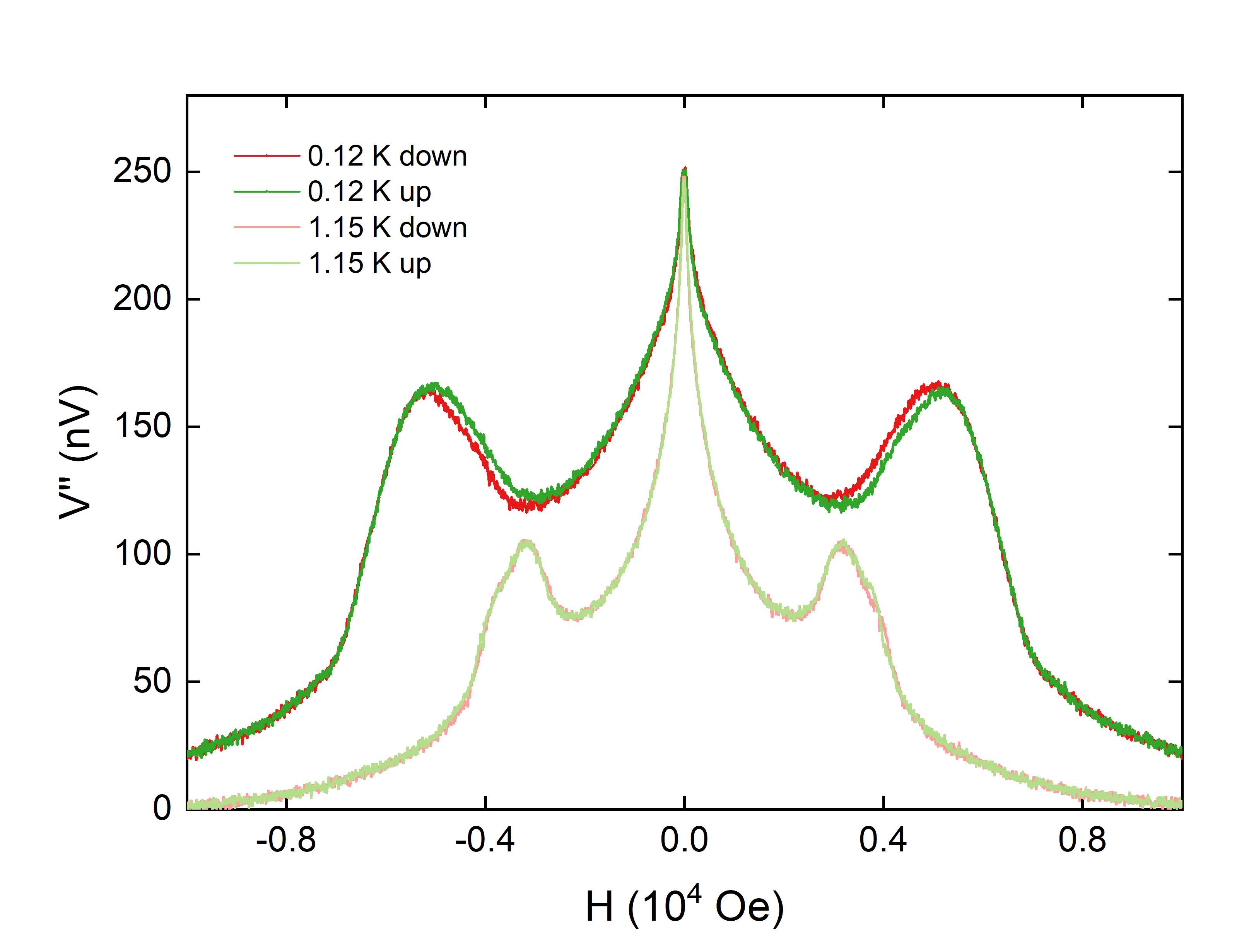}
		\label{figs4-1}
	\end{subfigure}
	\begin{subfigure}{}
		\includegraphics[width=0.48\columnwidth]{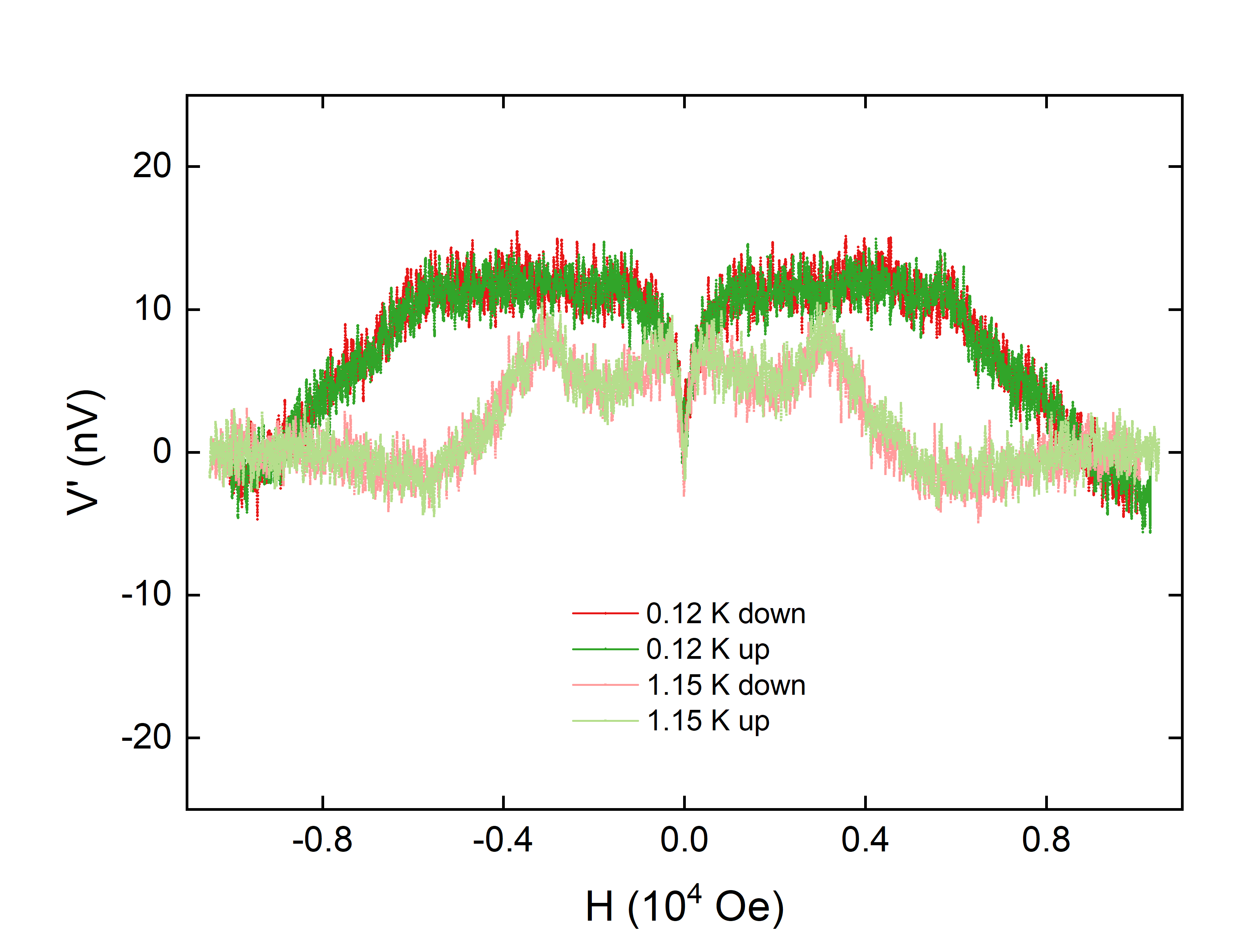}
		\label{figs4-2}
	\end{subfigure}
	\begin{subfigure}{}
		\includegraphics[width=0.48\columnwidth]{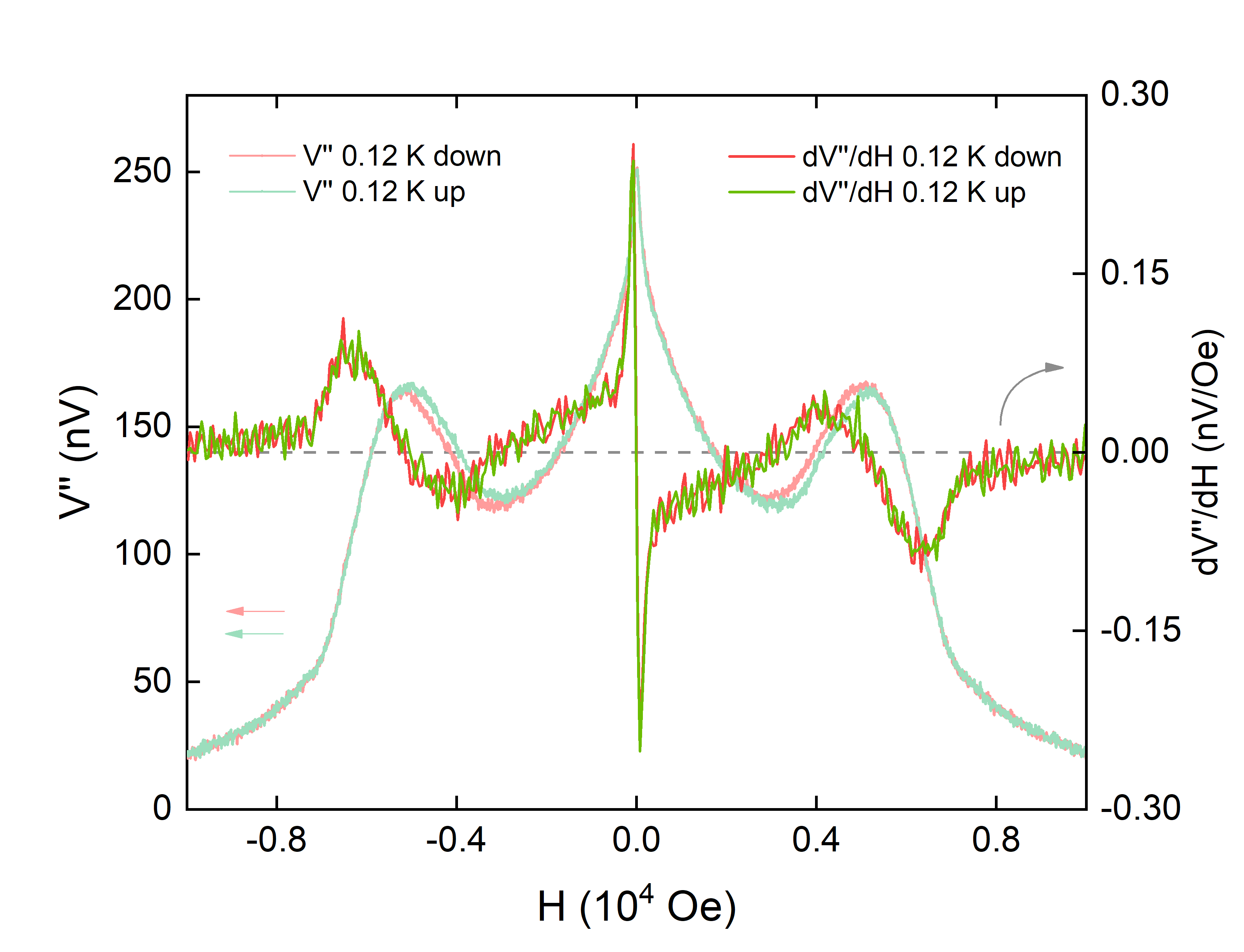}
		\label{figs4-3}
	\end{subfigure}
	\begin{subfigure}{}
		\includegraphics[width=0.48\columnwidth]{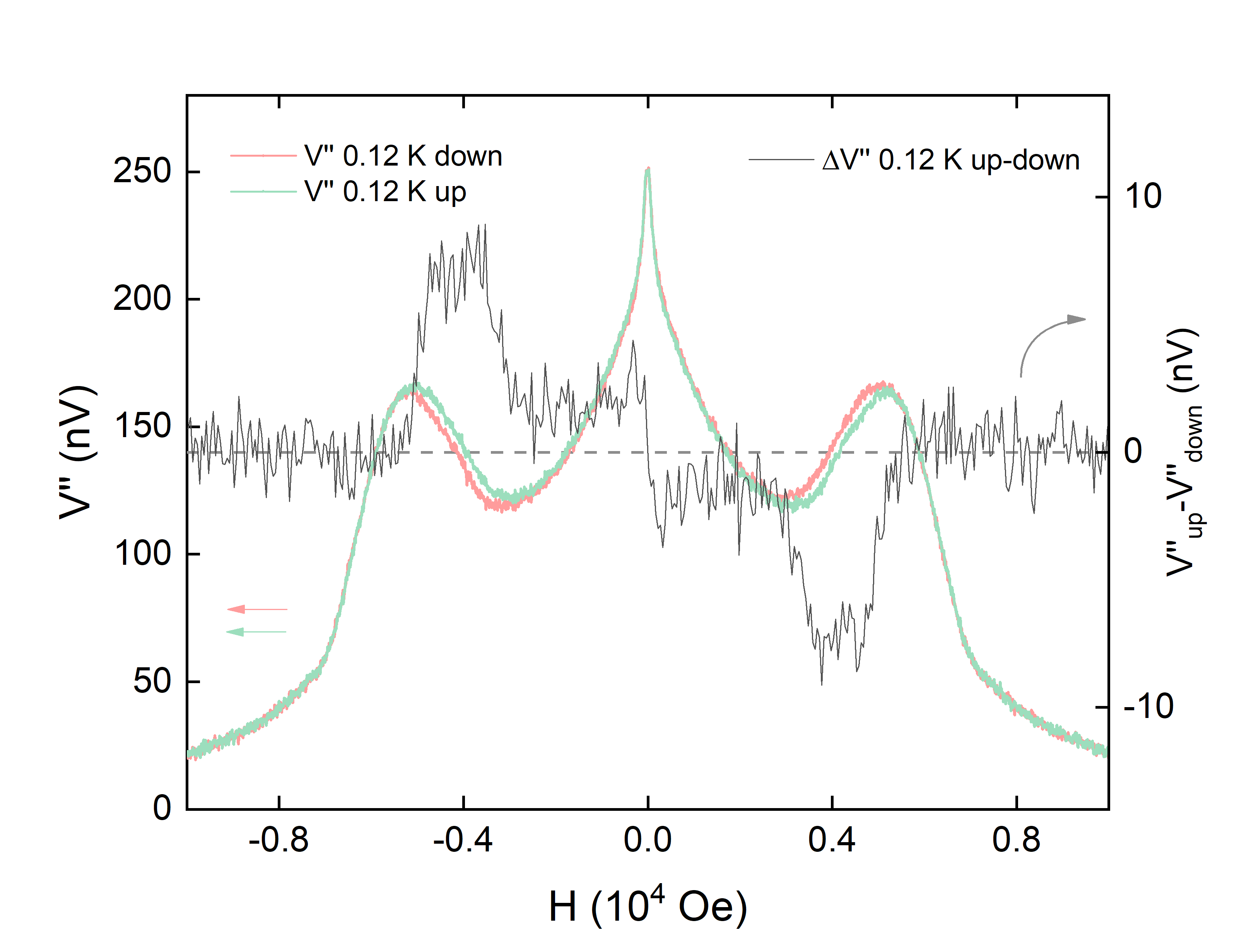}
		\label{figs4-4}
	\end{subfigure}
	\caption{Inductive (screening; upper left) and dissipative (upper right) response of CsV$_3$Sb$_5$ as a function of external magnetic field at different temperatures. Data taken with the same drive current as in the main text Fig.~\ref{fig3} at 100 kHz. (Lower left) First order derivative $dV''/dH$ at 0.12 K, overlaid with $V''(H)$. (Lower right) Hysteresis split $\Delta V''_\text{u-d}$ at 0.12 K, overlaid with $V''(H)$.}
\end{figure}
To improve data quality and to reduce the effect of finite time constant in lock-in detection, we repeated the field-dependent measurement at a much lower field ramp rate of $\sim$ 1 Oe/s in both ramp directions. For reference, the feature size relevant to the peak effect is $\sim 1000$ Oe. This allows us to calculate the first order derivative versus magnetic field $dV''/dH$. The local maxima/minima can be pinpointed whereas a peak in $dV''/dH$ at $\sim\pm 6.5$ kOe corresponds to the melting transition from strong to weak pinning.

As seen in Fig.~S4, the resulting data is highly symmetric about zero field at large fields. The hysteresis in the leading edges of the peaks are well resolved and consistent in both directions of applied field. Beyond that, we also notice a weaker hysteresis within $\pm 0.3$ kOe, corresponding to the Bragg glass phase, although it is barely visible in the $V''(H)$ data.

\newpage
\section{V. Temperature- and field-dependence of $V''$ at 20 $\text{kHz}$}
\begin{figure}[h]
	\centering
	\begin{subfigure}{}
		\includegraphics[width=0.48\columnwidth]{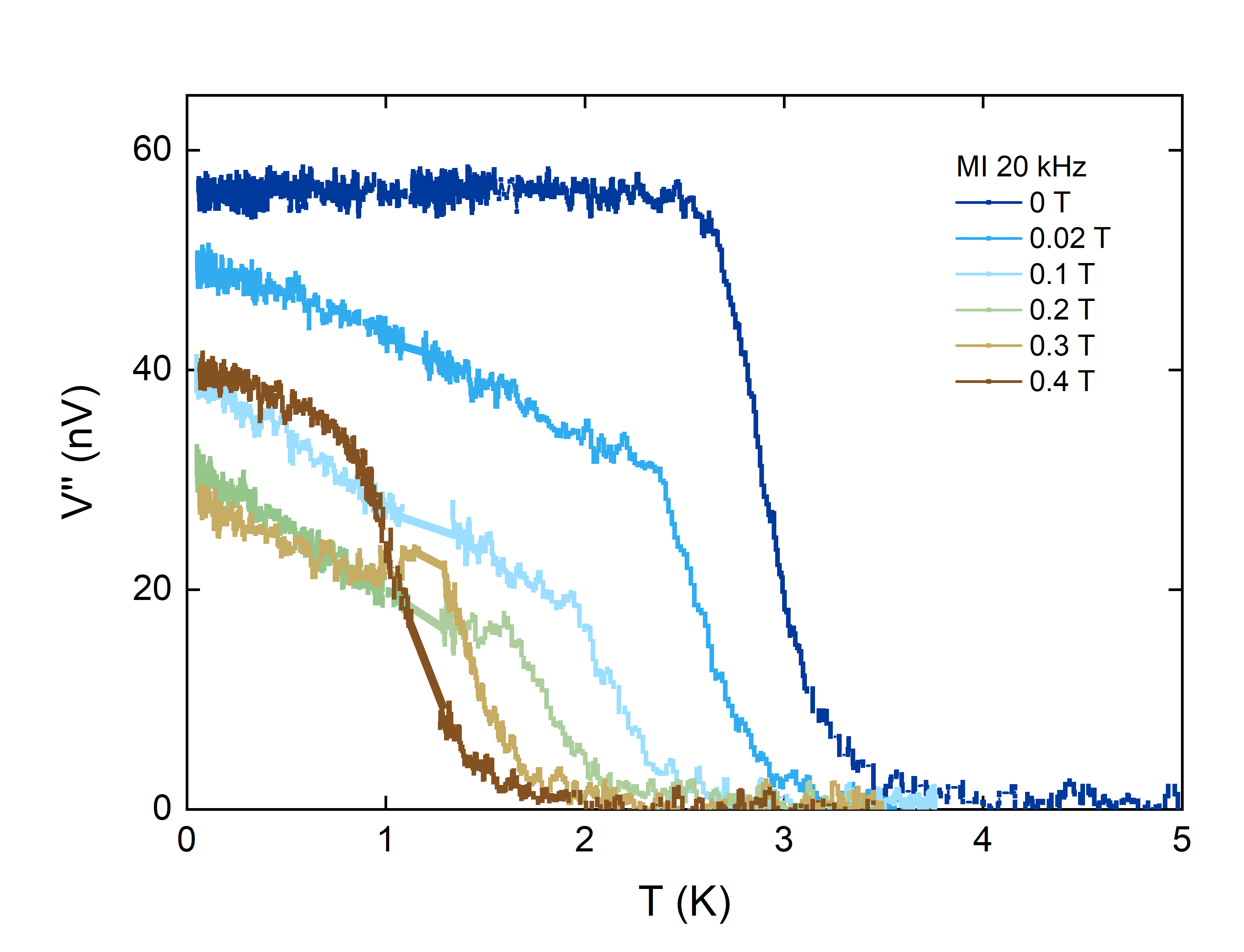}
		\label{figs5-1}
	\end{subfigure}
	\begin{subfigure}{}
		\includegraphics[width=0.48\columnwidth]{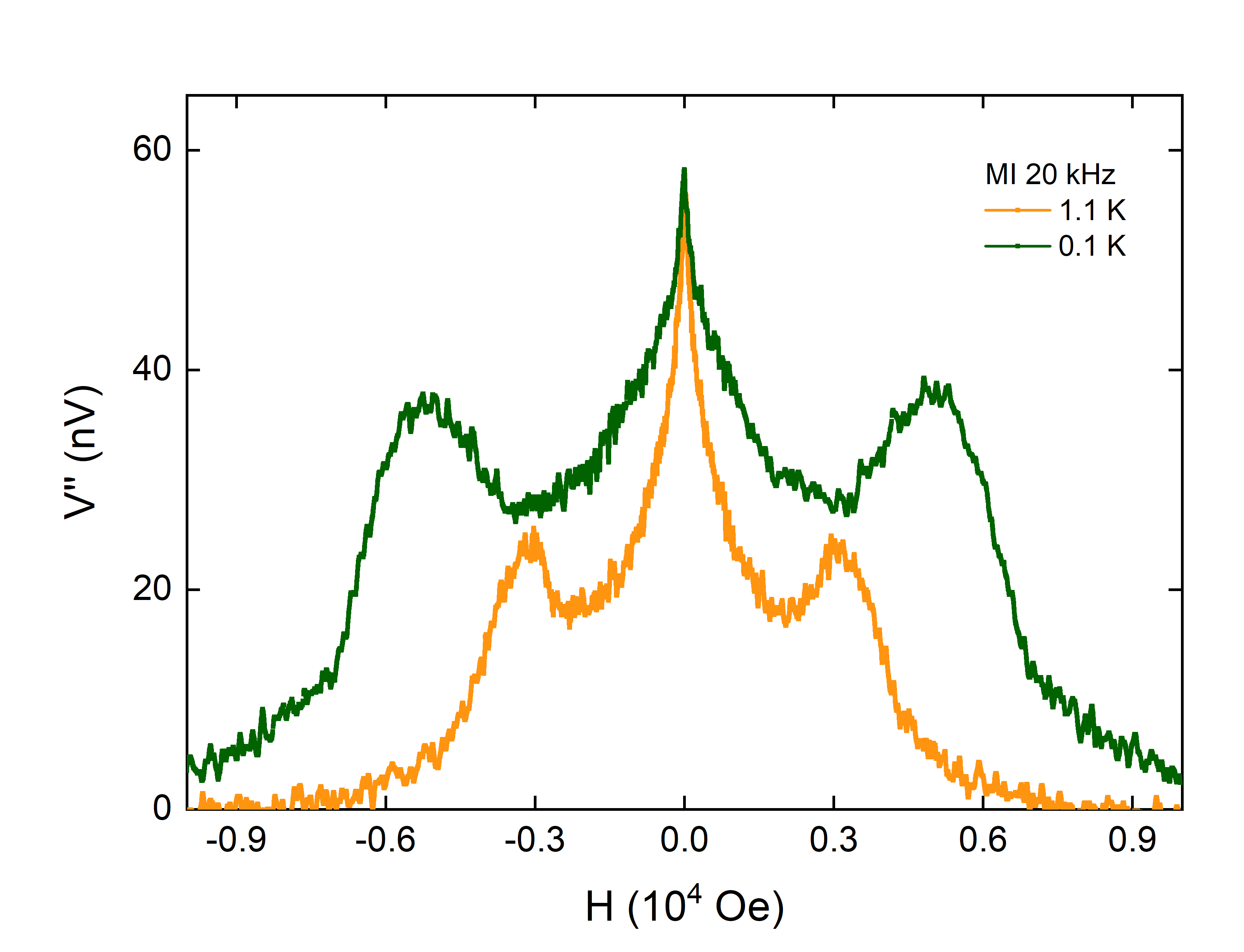}
		\label{figs5-2}
	\end{subfigure}
	\caption{Temperature- (left) and field-dependence (right) of inductive (screening) response of CsV$_3$Sb$_5$ at 20 kHz.}
\end{figure}
The temperature- and field-dependent screening response measurement is repeated at a drive frequency of 20 kHz at the same drive current. Since mutual inductance is proportional to both frequency and drive current, the signal amplitude is scaled down by a factor of 5. Nevertheless, we found the same non-monotonic field-dependence in screening response, consistent with the peak effect. 

Following the same method used in the 100-kHz analysis, we extracted the features $T_\text{knee}$, $T_\text{break}$, $T_c$, $H_\text{onset}$, $H_\text{peak}$, $H_\text{base}$, and $H_\text{irr}$. The extracted features are plotted in Fig. 4 in the main text. While features relevant to the peak does not differ as frequency changes, the irreversibility line (\textit{i.e.} onset of screening response) does shift to a lower temperature, as expected from a frequency-dependent activation energy for vortex depinning.

\newpage
\section{VI. Resistance and current-voltage characteristics of C$\text{s}$V$_3$S$\text{b}_5$}
\begin{figure}[h]
	\centering
	\begin{subfigure}{}
		\includegraphics[width=0.48\columnwidth]{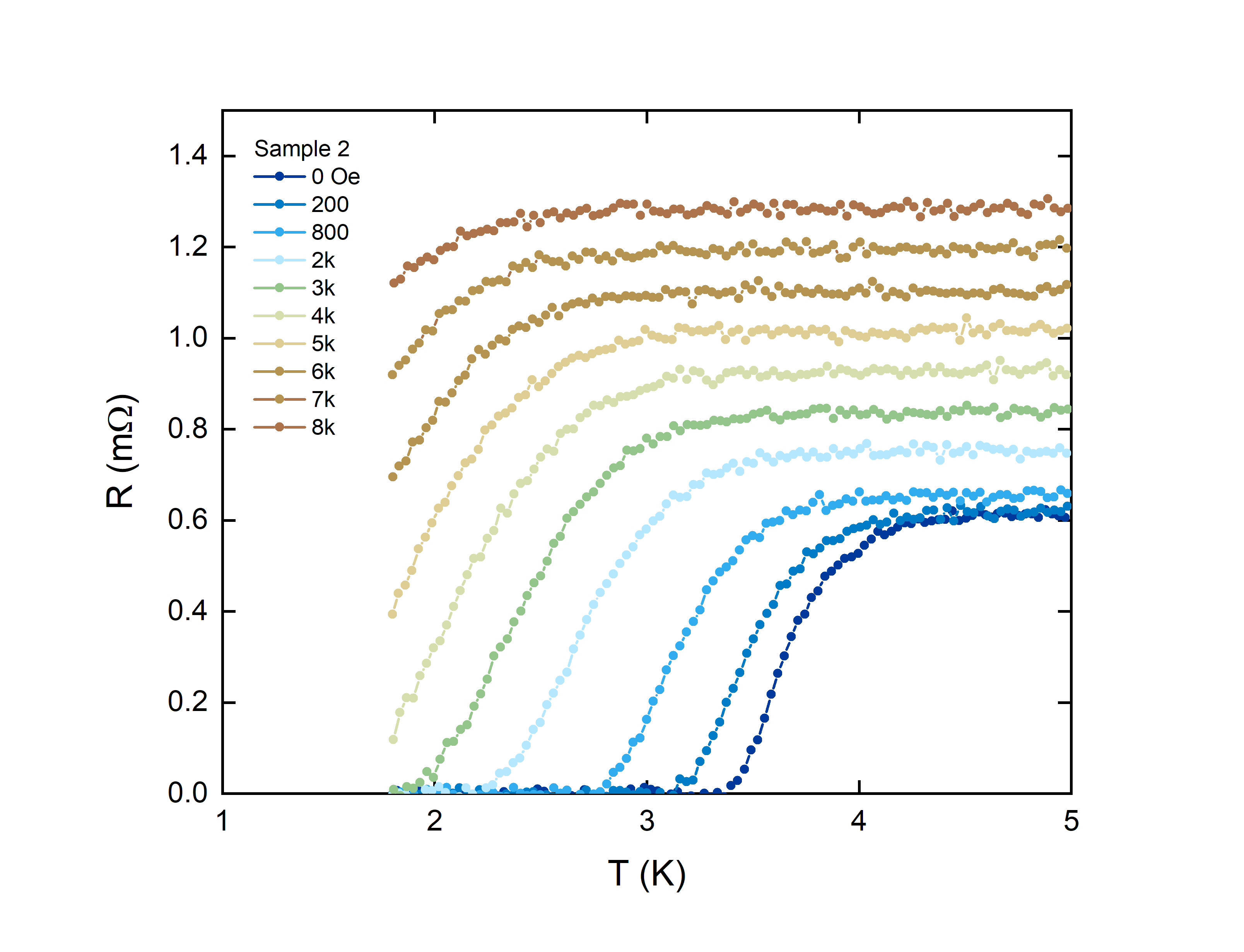}
		\label{figs6-1}
	\end{subfigure}
	\begin{subfigure}{}
		\includegraphics[width=0.48\columnwidth]{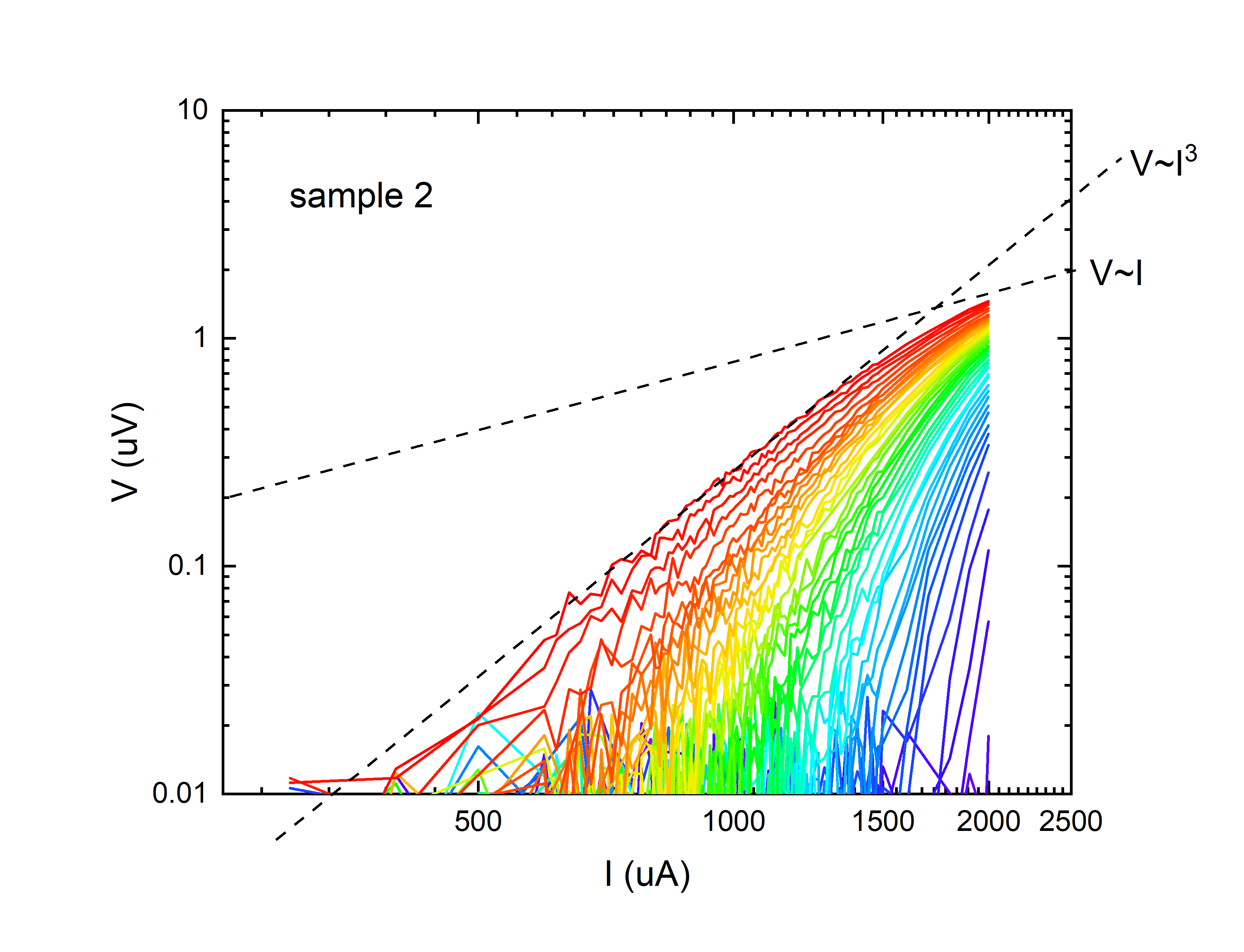}
		\label{figs6-2}
	\end{subfigure}
	\caption{Resistance vs. temperature in different magnetic fields (left) and I-V characteristics at 1.88 K (right) of CsV$_3$Sb$_5$ (Sample 2). Applied magnetic field ranges from 0 Oe (violet) to 3 kOe (red).}
\end{figure}
Resistance vs. temperature is measured in a Quantum Design Physical Properties Measurement System (PPMS). The sample for this measurement is an approximately $2\times0.5\times0.1$ mm$^3$ CsV$_3$Sb$_5$ crystal (Sample 2) within the same batch of samples measured by MI. Contacts were made with silver paste in a four-probe geometry. The excitation current we used for resistance measurement is 300 $\mu$A.

At zero applied magnetic field, the zero-resistance superconducting transition temperature $T_c\approx 3.25$ K, consistent with the onset of superfluid screening response $V''$ in MI data when extrapolated to $f\to0$ (dc). The field-dependent $T_c(H)$ is taken as the onset of superconducting transition, where the resistance starts to deviate from its normal state value by one standard deviation of the signal fluctuation. The broadened superconducting transition suggests the crucial role of vortex dynamics in this material.

Indeed, the I-V characteristics at $T=1.88$ K in varying magnetic fields show a strongly non-Ohmic behavior over a broad range of excitation currents, further supporting the vortex glass picture. The applied magnetic field ranges from 0 Oe (violet) to 3 kOe (red). From the I-V characteristics we extract the critical current by vortex depinning.

\newpage
\section{VII. Peak effect in critical current at 1.88 K}
\begin{figure}[h]
	\centering
	\includegraphics[width=0.6\columnwidth]{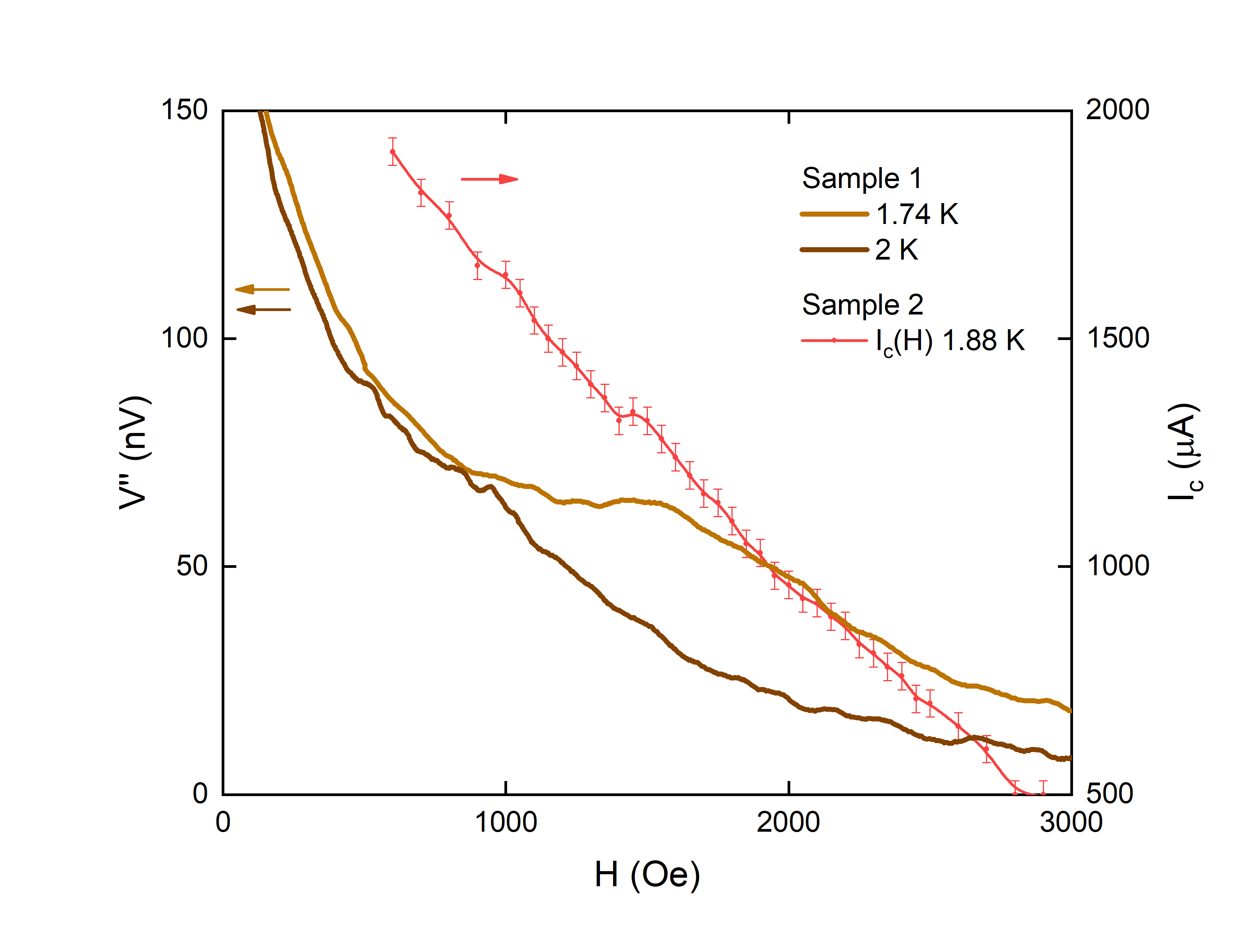}
	\label{figs7}
	\caption{Critical current as a function of external magnetic field, compared with screening response at similar temperatures.}
\end{figure}
The critical current extracted from the I-V characteristics of Sample 2 (right axis) is compared with screening response $V''(H)$ of Sample 1 (left axis). Despite a small one, the peak in critical current is seen at similar $H_\text{peak}$ as the peaks in $V''(H)$. As peak effect is considered originating from enhancement in pinning force density $f_p$, and thus critical current $j_c=f_p/B$, this is a solid evidence for the observation of the peak effect in CsV$_3$Sb$_5$.

\newpage
\section{VIII. Temperature-dependence of dc magnetization}
\begin{figure}[h]
	\centering
	\includegraphics[width=\columnwidth]{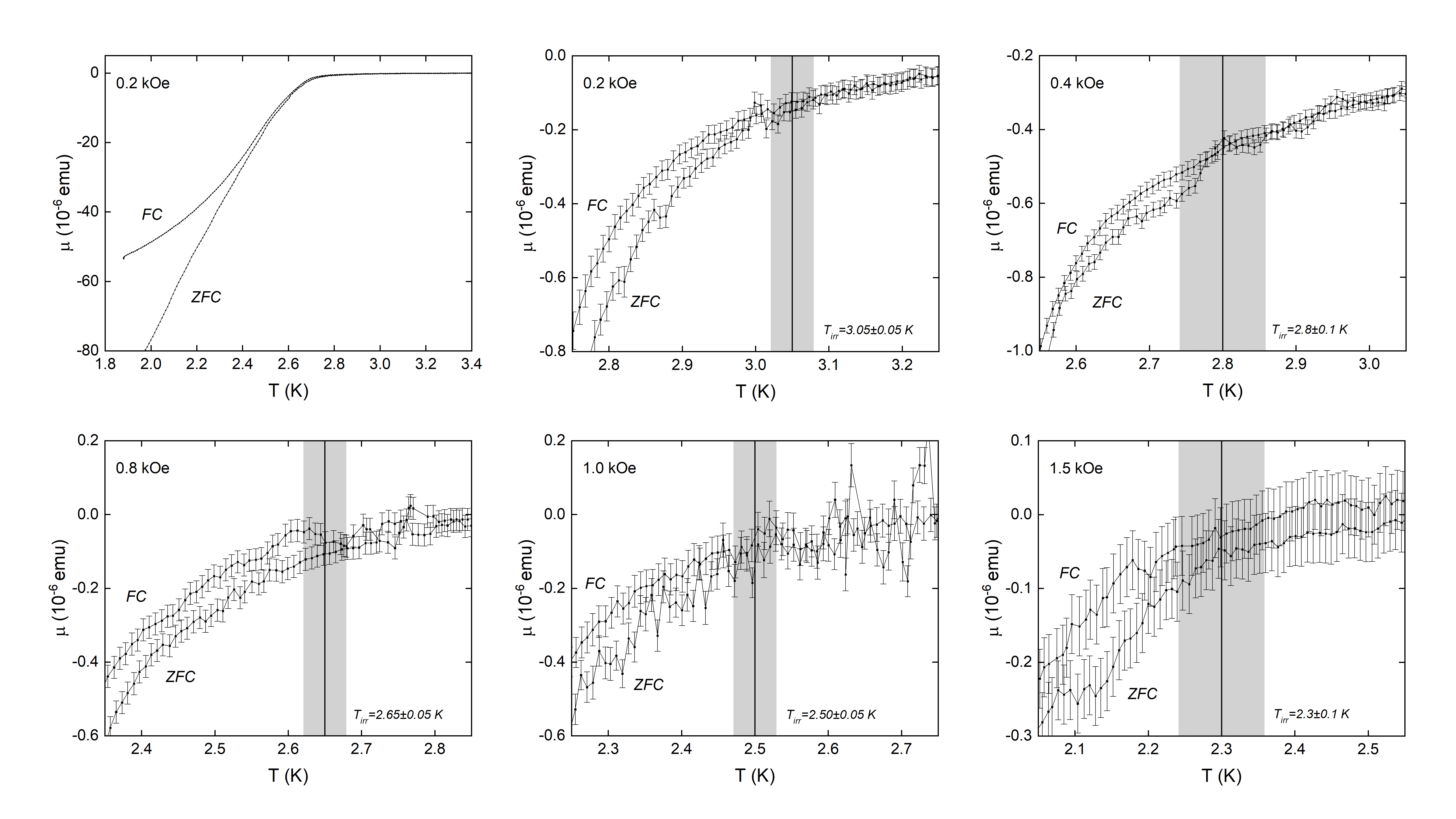}
	\label{figs8}
	\caption{Critical current as a function of external magnetic field, compared with screening response at similar temperatures.}
\end{figure}
The dc magnetization of another CsV$_3$Sb$_5$ sample (approximately $1.5\times1.5\times0.1$ mm$^3$) (Sample 3) within the same batch was measured using a superconducting quantum interference device (SQUID) magnetometer in a Quantum Design Magnetic Properties Measurement System (MPMS). The sample was mounted using GE 7031 varnish such that the magnetic field is along the c-axis.

Magnetization was measured by the standard ZFC/FC protocol: as an example in the first panel in Fig.~S8, starting from 6 K where no superconductivity is expected, the sample was cooled in zero field (remnant field $\sim 0.5$ Oe); At 1.8 K, we ramp up the c-axis magnetic field to a desired strength. ZFC data were collected during the subsequent warming to 6 K, where the full diamagnetic response relaxes to zero. Immediately following was a cooling step in the same magnetic field, during which FC data were collected as the field-trained diamagnetic response. The splitting of ZFC/FC curves was taken as the onset of irreversibility in flux pinning, and thus termed the irreversibility line $T_\text{irr}(H)$.

The next five panels in Fig.~S8 show the evolution of the irreversibility line $T_\text{irr}(H)$. Here, the splitting of ZFC/FC occurs gradually and generally precedes the rapid increase in diamagnetic response. Therefore, we need to zoom in to very closely to zero signal and consequently the determination of the splitting suffers from signal statistical fluctuations. Nevertheless, with a generous error bar, we extract $T_\text{irr}(H)$ for a handful of applied field, which were included in the vortex phase diagram Fig.~\ref{fig4} in the main text.
\end{document}